\documentclass[a4paper]{JHEP3}
\usepackage{epsfig}

\title{Joint resummation for pion wave function and pion  transition form factor}

\author{Hsiang-nan Li$^{a}$, Yue-Long Shen$^{b}$,  Yu-Ming Wang$^{c}$
\footnote{Aachen: TTK-13-22, SFB/CPP-13-75; Munich: TUM-HEP-902/13. }

{\it \small \hspace{0.3 cm} $^a$Institute of Physics, Academia Sinica,
Taipei, Taiwan 115, Republic of China; \\
Department of Physics, National Cheng-Kung University, Tainan, Taiwan 701,
Republic of China; \\
Department of Physics, National
Tsing-Hua University, Hsinchu, Taiwan 300, Republic of China}
\\
{\it \small $^b$College of Information Science and Engineering,
Ocean University of China, Qingdao, Shandong 266100, P.R. China}
\\
{\it \small $^c$ Institut f\"{u}r Theoretische Teilchenphysik und Kosmologie
RWTH Aachen, D-52056 Aachen, Germany;
\\
Physik Department T31, James-Franck-Stra${\ss}$e,
Technische Universit\"{a}t M\"{u}nchen, D-85748 Garching, Germany}

}

\abstract{We construct an evolution equation for the pion  wave function
in the $k_T$ factorization theorem, whose solution sums the mixed
logarithm $\ln x\ln k_T$ to all orders, with $x$ ($k_T$) being a
parton momentum fraction (transverse momentum). This joint resummation
induces strong suppression of the pion wave function in the small $x$
and large $b$ regions, $b$ being the impact parameter conjugate to $k_T$,
and improves the applicability of perturbative QCD to hard exclusive processes.
The above effect is similar to those from the conventional threshold
resummation for the double logarithm $\ln^2 x$ and the conventional
$k_T$ resummation for $\ln^2 k_T$. Combining the evolution equation for
the hard kernel, we are able to organize all large logarithms in the
$\gamma^{\ast} \pi^{0} \to \gamma$ scattering, and to establish a
scheme-independent $k_T$ factorization formula. It will be shown that the
significance of next-to-leading-order contributions and saturation behaviors
of this process at high energy differ
from those under the conventional resummations. It implies that
QCD logarithmic corrections to a process must be handled
appropriately, before its data are used to extract a hadron wave
function. Our predictions for the involved pion transition form factor,
derived under the joint resummation and the input of a non-asymptotic pion
wave function with the second Gegenbauer moment $a_2=0.05$, match
reasonably well the CLEO, BaBar, and Belle data.
}


\begin{document}


\section{INTRODUCTION}

Great efforts have been devoted to the extension of the $k_T$ factorization theorem
for exclusive processes \cite{CCH,CE,LRS,BS,LS,HS} to subleading levels
recently. The next-to-leading-order (NLO) corrections to the pion transition
(electromagnetic) form factor associated with the
$\pi\gamma^*\to \gamma(\pi)$ scattering have been calculated at leading power
\cite{Nandi:2007qx,LSW11}. Those to the $B\to\pi$ transition
form factors involved in $B$ meson semileptonic decays were derived in
\cite{LSW12}. Up to subleading power, the three-parton
contributions to the pion electromagnetic form factor, to the $B\to\gamma$
transition form factor, and to the $B\to\pi$ transition form factors
have been studied in  \cite{CL1104}, \cite{Charng:2005fj}, and \cite{CL1112},
respectively. A $k_T$-dependent hard kernel is defined as the difference
between QCD diagrams and effective diagrams for transverse-momentum-dependent
(TMD) hadron wave functions. Therefore, to obtain a NLO hard kernel, both
QCD diagrams and effective diagrams need to be evaluated at the same level.
The NLO analysis of the $B$ meson and pion wave functions have revealed
various important logarithms, which stimulate corresponding resummation formalisms
for their organization to all orders in the strong coupling constant.

A TMD hadron wave function
contains the light-cone singularity from the region with a loop momentum
parallel to a Wilson line on the light cone \cite{Co03}. To regularize the
light-cone singularity, one may rotate the Wilson line away from the
light cone to an arbitrary direction $u$ with $u^2\not=0$ \cite{Co03,CS91}.
The higher-order wave function then depends on $u^2$ through the scale
$\zeta_P^2 \equiv 4(P \cdot u)^2/u^2$, where $P$ denotes the hadron
momentum. The variation of $u$, namely, of $\zeta_P^2$ introduces a
factorization-scheme dependence into the hadron wave function. The evaluation
of the NLO effective diagrams for the $B$ meson wave
function indicates the existence of the logarithms $\ln^2(\zeta_P^2/m_B^2)$
and $\ln x\ln(\zeta_P^2/m_B^2)$ \cite{LSW12}, $m_B$ being the $B$ meson mass
and $x$ being the momentum fraction of the spectator. The NLO diagrams for
the pion wave function produce the mixed logarithm $\ln x\ln(\zeta_P^2/k_T^2)$
\cite{LSW11}, $k_T$ being the parton transverse momentum. All the
above logarithms become large as $u^2\to 0$, and as $x$ and $k_T$ are small,
which is the dominant kinematic region in the $k_T$ factorization theorem for
exclusive processes. The logarithms in the $B$ meson wave function have been
organized under the rapidity resummation \cite{Li:2012md}, whose
effect was shown to diminish the $B$ meson wave function at the end point $x\to 0$.

The above observation hints that the resummation of the mixed logarithm
$\ln x\ln(\zeta_P^2/k_T^2)$ for the pion wave function would modify both
the $x$ and $k_T$ dependencies. It then calls for the joint resummation
\cite{LL99,Li99,LSV00,Bozzi07}, which was proposed to unify the
conventional threshold resummation for $\ln^2x$ \cite{S,CT,KM} and the
conventional $k_T$ resummation for $\ln^2 k_T$ \cite{CS91,Collins:1984kg}.
For a recent review on this subject, see \cite{Li:2013ela}.
In this paper we will construct an evolution equation in the scale $\zeta_P^2$
following the idea in \cite{Li99}, whose solution resums the mixed logarithm
in the Mellin ($N$, conjugate to $x$) and impact-parameter ($b$, conjugate
to $k_T$) spaces. The inverse Mellin transformation is then applied to get
the $x$ dependence of the pion wave function. It will be demonstrated that
the joint resummation induces suppression which is stronger at small $x$ than at
moderate $x$, and intensifies with increase of
$b$. This effect, similar to those of the threshold and $k_T$
resummations, improves the applicability of perturbative QCD (PQCD) to hard exclusive
processes. Combining the evolution equation for the hard kernel of the
$\gamma^{\ast} \pi^{0} \to \gamma$ scattering, we organize all the relevant
large logarithms, and remove the factorization-scheme dependence on $\zeta_P^2$
mentioned before. This is the first time that the $k_T$ factorization for a
simple exclusive process can be made scheme independent in the presence of
the light-cone singularity.

It has been known that $\gamma^{\ast} \pi^{0} \to \gamma$ serves
as an ideal process for the determination of the pion wave function, and
the involved pion transition form factor $F(Q^2)$, $Q^2$ being the
momentum transfer squared, has been investigated thoroughly. In particular,
it was claimed that the quantity $Q^2F(Q^2)$ (including those for the
$\eta$ and $\eta'$ meson transition form factors) begins to
saturate at relatively low $Q^2$ as calculated
in the hard scattering approach \cite{Brodsky:2011yv},
in QCD sum rules (QCDSR) \cite{Melikhov},
in light-cone sum rules (LCSR) \cite{Stefanis},
and in the light-front holographic QCD \cite{Brodsky:2011xx}.
We will analyze the leading-order (LO) and NLO contributions to the pion transition
form factor with inputs of different model wave functions, including
the asymptotic model, the flat model, and the model with the second Gegenbauer
moment $a_2$. The results are compared to those from the PQCD approach
\cite{LM09}, that incorporates the conventional threshold and $k_T$ resummations.
It will be observed that the significance of the NLO correction to
and the saturation behavior of $Q^2F(Q^2)$ differ under the joint
resummation and the conventional resummations. It implies that
QCD logarithmic corrections to a process must be handled
appropriately, before its data are used to extract a hadron wave
function. Our predictions for $Q^2F(Q^2)$ from a non-asymptotic
pion wave function with $a_2=0.05$ match reasonably well the CLEO, BaBar,
and Belle data, which seem to indicate scaling violation at currently
accessible $Q^2$.

In Sec.~\ref{section: evolution equation} we construct the evolution equation
for the resummation of the mixed logarithm in the pion wave function, and then
solve it in the Mellin and impact-parameter spaces. The inverse Mellin
transformation of the solution is performed in
Sec.~\ref{section: resummation improved wavefunctions}, with different
initial conditions of the evolution. Note that the running
of the strong coupling constant down to the low energy region has to be
modified in order to avoid the Landau pole. The joint resummation effect
on the $x$ and $b$ dependencies of the pion wave function is then examined.
In Sec.~\ref{section: pion transition form factor} the pion transition
form factor is evaluated for a given model wave function at the LO and NLO
levels under the joint resummation and the conventional resummations. The
different outcomes for the NLO contributions and for the saturation
behaviors at high energy are compared.
We summarize our findings, and discuss potential extension of our formalism to
more complicated processes in Sec.~\ref{section: conclusion}. The explicit
expressions for the solutions of the evolution equation are collected in
Appendix~\ref{functions F}.

\section{EVOLUTION EQUATION}
\label{section: evolution equation}

The TMD pion wave function $\Phi(x,k_{T})$ is defined by the
non-local hadron-to-vacuum matrix element
\footnote{The leading-twist light-cone projector for a pion
in the collinear factorization can be found in \cite{Beneke:2000wa}.}
\begin{eqnarray}
\Phi(x,k_{T}, \zeta^2, \mu_f)&=&\int\frac{dy^+}{2\pi
}\frac{d^2y_T}{(2\pi)^2}e^{-i x P^- y^{+} +i {\bf k}_{T}\cdot {\bf
y}_T} \nonumber \\
&& \times\langle 0 |{\bar q}(y) W_y(u)^{\dag} \, I_{u;y,0} \, W_0(u) \not
n_+\gamma_5 q(0)| \pi (P)  \rangle\,,
\label{de1}
\end{eqnarray}
where $\mu_f$ is the factorization scale,
the coordinate $y=(y^+,0,{\bf y}_{T})$ is off
the light cone generally, and $x P^-$ and ${\bf k}_{T}$ are the longitudinal
and transverse momenta carried by the anti-quark $\bar q$, respectively. A TMD hadron wave function
describes the distributions of a light parton in both light-ray and transverse directions.
To maintain the gauge invariance of  the definition in Eq. (\ref{de1}), the gauge-link operator $W_y(u)$
\begin{eqnarray}
\label{eq:WL.def}
W_y(u) = {\cal P} \exp\left[-ig \int_0^\infty \, d\lambda \,
u \cdot A(y+\lambda u)\right],
\end{eqnarray}
has been introduced, where $g$ is the QCD coupling constant, and $\cal P$ denotes the path-ordered exponential.
The  non-light-like vector $u$, different from the usual Wilson line direction $n_+=(1,0,{\bf 0}_T)$,
plays a role of the regulator for the light-cone divergences \cite{Co03}.
The transverse gauge link $I_{u;y,0}$, unraveling the cusp obstruction in the contour of the Wilson
lines at infinity, does not  contribute  in the covariant gauge \cite{CS08}.

As determining a NLO hard kernel in the $k_T$ factorization formula for a pion-induced process, we perform
the infrared substraction defined as the convolution of the NLO pion wave function
with the LO hard kernel. The QCD correction to the pion wave
function gives rise to the mixed logarithm
$\ln x\,\ln(\zeta^2P^{-2}/k_T^2)$ \cite{Nandi:2007qx,LSW11,LSW12},
with the dimensionless rapidity parameter
\begin{eqnarray}
\zeta^2=\frac{4(n_{-} \cdot u)^2} {u^2} \,,
\label{rapidity parameter}
\end{eqnarray}
$n_-=(0,1,{\bf 0}_T)$ being a light-like vector along the pion momentum $P$.
The double rapidity logarithm $\ln^2 \zeta^2$ in the $B$ meson case
is absent here because of the color-transparency mechanism for an energetic
pion, which suppresses soft gluon contributions.

The goal of this section is to construct an evolution equation, whose solution sums the mixed
logarithm in the pion wave function. Following \cite{Li99},
we trade the derivative with respect to the rapidity parameter $\zeta^2$ for the variation of the Wilson
link direction $u$,
\begin{eqnarray}
\zeta^2{d \over d \zeta^2} \Phi=-{ u^2 \over n_{-} \cdot u}
{n_{-}^{\alpha} \over 2} {d \over d u^{\alpha}} \Phi \,.\label{chain}
\end{eqnarray}
It is obvious that this chain rule simplifies the analysis dramatically as the $u$
dependence appears only through the Wilson line interactions.
Applying Eq.~(\ref{chain}) to the Feynman rule associated with the Wilson link, we have
\begin{eqnarray}
\zeta^2{d \over d \zeta^2} {u^{\beta} \over u \cdot l+ i \epsilon}
={ \hat{u}^{\beta} \over 2 u \cdot l} \,,
\end{eqnarray}
with the special vertex
\begin{eqnarray}
\hat{u}^{\beta}={ u^2 \over n_{-} \cdot u} \left ( {n_{-} \cdot l \over u \cdot l} u^{\beta}
- n_{-}^{\beta}   \right ) \,.
\end{eqnarray}
We will derive the rapidity evolution equation
\begin{eqnarray}
\zeta^2{d \over d \zeta^2} \Phi(x,k_{T}, \zeta^2, \mu_f) = \Gamma(x,k_{T}, \zeta^2)
\otimes \Phi(x,k_{T}, \zeta^2, \mu_f) \,,
\label{evolution equation}
\end{eqnarray}
where $\otimes$ represents convolutions in the momentum fraction $x$ and
the transverse momentum $k_T$, and the evolution kernel $\Gamma$ involves the
diagrams with the special vertex.

\begin{figure}[ht]
\begin{center}
\hspace{-1 cm}
\includegraphics[scale=0.6]{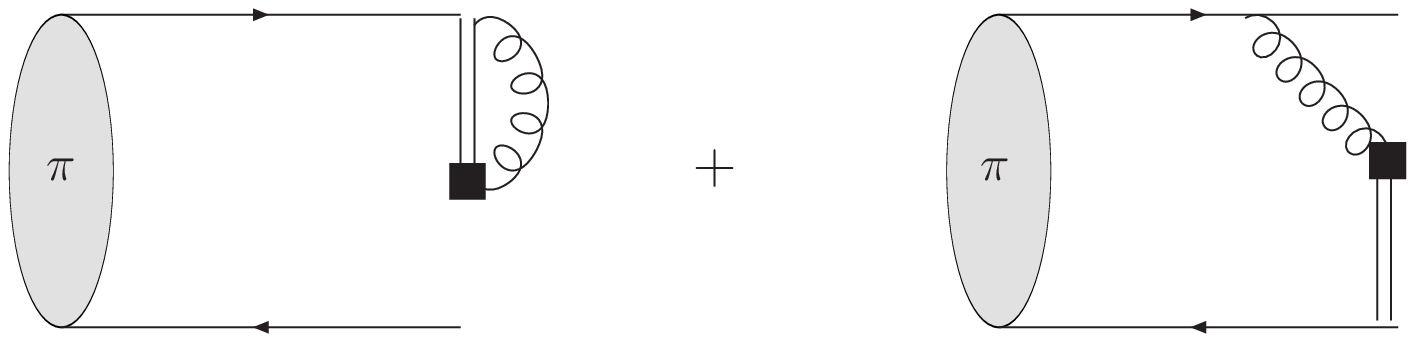}
\\(a)\\
\vspace{1 cm}
\includegraphics[scale=0.6]{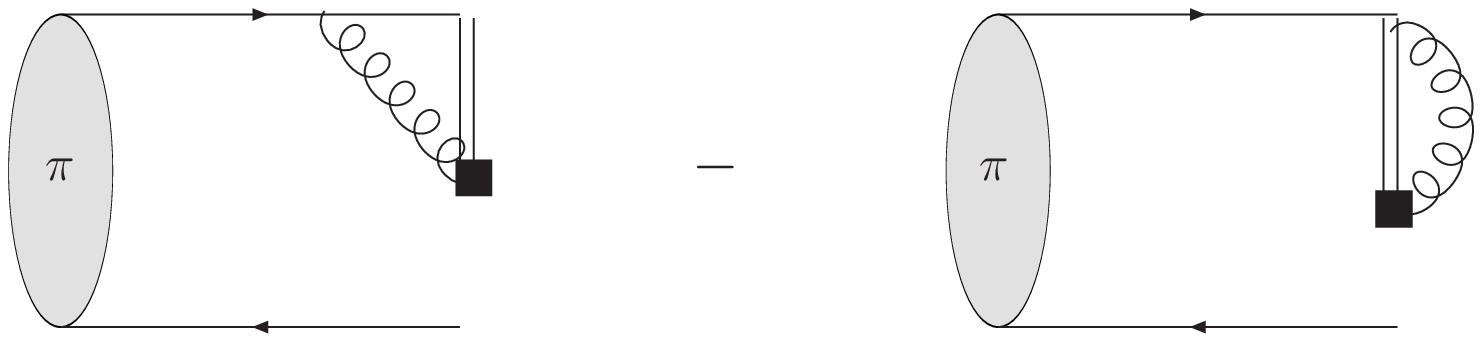}\\(b)
\caption{(a) diagrams for the function $K$ from soft gluon exchanges between the
Wilson lines and the active quark, and (b) diagrams for the function $G$ from
hard gluon exchanges, where the box denotes the special vertex.
The second diagram in the function $G$ is included to avoid the double counting
of the soft contribution.
Those diagrams with gluon radiations off the spectator quark are not displayed here. }
\label{K and G functions}
\end{center}
\end{figure}

\subsection{Evolution kernel}

It is easy to see that the structure of the special vertex
suppresses a collinear gluon contribution to $\Gamma$ \cite{Li:2012md}.
The evolution kernel is then dominated by soft and hard gluon exchanges,
usually denoted as the functions $K$ and $G$, respectively.
The soft and hard gluon radiations off the active quark,
as shown in Fig.~\ref{K and G functions}, lead to
\begin{eqnarray}
K_1&=&-{i g^2 C_F \over 2}\int \frac{d^4l}{(2\pi)^4}\frac{\hat u \cdot n_-}{(u \cdot l+i\epsilon)
(l^2+i\epsilon) (n_{-} \cdot l+i\epsilon)},\\
K_2 \otimes \Phi &=& {i g^2 C_F \over 2} \int \frac{d^4l}{(2\pi)^4}
\frac{\hat u \cdot n_-}{(u \cdot l+i\epsilon)(l^2+i\epsilon) (n_{-} \cdot l+i\epsilon)}
\nonumber \\
&& \times  \Phi(x-l^-/P^-,|{\bf k}_T-{\bf l}_T|,\zeta^2,\mu_f)\,,
\label{evolution kernel of active quark: K function}
\end{eqnarray}
for the function $K$, and
\begin{eqnarray}
G_1&=&-{i g^2 C_F \over 2}\int \frac{d^4l}{(2\pi)^4}\frac{ (\bar x \not \! P+ \not  l) \not \! \hat{u} }
{(u \cdot l+i\epsilon) (l^2+i\epsilon) [(\bar x P +l)^2 +i\epsilon]} \,,    \nonumber \\
G_2&=&K_1  \,,
\label{evolutionkernel of active quark: G function}
\end{eqnarray}
for the function $G$ with the variable $\bar{x}\equiv1-x$.

Adopting the dimensional regularization for the ultraviolet divergence and
regularizing the infrared divergence with the gluon mass $\lambda$, we obtain
\begin{eqnarray}
K_1=-{\alpha_s \,  C_F \over 4 \pi}\left ( {1 \over \epsilon} - \gamma_E + \ln {4 \pi \mu^2 \over \lambda^2 }\right )\,.
\label{K1: result}
\end{eqnarray}
Since the soft divergences cancel between $K_1$
and $K_2$, and between $G_1$ and $G_2$, the gluon mass $\lambda$ will
approach to zero eventually. For the evaluation of $K_2$,
we apply the Mellin and Fourier transformations
\begin{eqnarray}
\tilde \Phi(N,b,\zeta^2,\mu_f)=\int_0^1 dx (1-x)^{N-1}\int \frac{d^2 k_T}{(2\pi)^2}
\, \, \exp(i{\bf k}_T\cdot {\bf b}) \, \Phi(x,k_T,\zeta^2,\mu_f)\,,
\label{Mellin and Fourier transformation}
\end{eqnarray}
$b$ being the impact parameter.
Equation~(\ref{evolution kernel of active quark: K function}) then gives
$\tilde K_2\, \tilde \Phi(N,b,\zeta^2,\mu_f)$ with the soft kernel
\begin{eqnarray}
\tilde K_2&=& {ig^2C_F \over 2} \int
\frac{d^4l}{(2\pi)^4}\left(1-\frac{l^-}{P^-}\right)^{N-1}
\exp(-i{\bf l}_T\cdot {\bf b}) \, \frac{ \hat u \cdot n_{-}}{(u \cdot
l+i\epsilon)(l^2+i\epsilon) (n_{-}\cdot l+i\epsilon)}\,, \nonumber\\
&=&{\alpha_s \, C_F \over 2 \pi} \,  \left [  K_0
\left(\lambda b \right) -  K_0\left(
\frac{\zeta P^{-} b}{N} \right) \right ],
\label{K2: Mellin and Fourier spaces}
\end{eqnarray}
in which the terms suppressed by powers of $1/\zeta^2$ have been dropped, and
$K_0$ is the zeroth-order modified Bessel function of the second kind.
Hence, the bare function $\tilde{K}^{(b)}$ is written as
\begin{eqnarray}
\tilde{K}^{(b)}=K_1+\tilde{K}_2=-{ \alpha_s \, C_F \over 4 \pi} \left ( {1 \over \epsilon} - \gamma_E
+ \ln { 4\pi \mu^2  \, N^2 \over \zeta^2 \,  P^{-2} }   \right )\,,
\label{bare K function: active quark}
\end{eqnarray}
where the large-$N$ expansion of Eq.~(\ref{K2: Mellin and Fourier spaces}) has been
made, and the superscript $(b)$ labels the bare function explicitly.

The bare hard function $G^{(b)}$ can be calculated following the same line, and reads
\begin{eqnarray}
G^{(b)}=G_1-G_2={\alpha_s \, C_F \over 4 \pi}\left [ {1 \over \epsilon}
- \gamma_E + \ln {4 \pi \mu^2 \over \zeta^2(\bar{x} P^{-})^2 }
- 4 \right ]\,.
\label{bare G function: active quark}
\end{eqnarray}
We will adopt the approximation $\bar{x} \approx 1$ in the small $x$ region, where
the mixed logarithm plays a significant role.
It is found that both the soft and hard functions depend on the factorization scale $\mu$,
and such a dependence cancels in their sum. This fact is attributed to
the $\mu$ independence of the mixed logarithm that we are going to resum.

Applying the modified minimal substraction ($\overline{\rm MS}$) scheme
to the ultraviolet renormalization yields
\begin{eqnarray}
\tilde{K}^{(r)}(\mu)&=&-{\alpha_s \, C_F \over 2 \pi}
\ln { \mu \, N  \over \zeta P^{-} } \,,   \qquad  \hspace{0.8 cm}
\lambda_{\tilde{K}}=\mu \frac{d \delta K}{ d \mu}={\alpha_s \, C_F \over 2 \pi} \,,
\nonumber \\
G^{(r)}(\mu)&=&{\alpha_s \, C_F \over 2 \pi}
\left (\ln { \mu \over \zeta P^{-} }  -2 \right ),  \qquad
\lambda_G= \mu\frac{d \delta G}{ d \mu}=-\lambda_{\tilde{K}} \,,
\end{eqnarray}
where the additive counterterms $\delta K$ ($\delta G$) of the
function $\tilde{K}^{(b)}$ ($G^{(b)}$)
can be read from  Eq.~(\ref{bare K function: active quark})
(Eq.~(\ref{bare G function: active quark})).
The renormalization-group (RG) equations for the soft and hard functions are then given,
in terms of the anomalous dimensions $\lambda_{\tilde{K}}$ and $\lambda_{G}$,  by
\begin{eqnarray}
\mu\frac{d \tilde{K}^{(r)}}{ d \mu}=-\lambda_{\tilde{K}}, \qquad
\mu\frac{d  G^{(r)}}{ d \mu}=-\lambda_{G} \,,
\end{eqnarray}
which lead to the RG improved evolution kernel
\begin{eqnarray}
\tilde{K}^{(r)}(\mu)+G^{(r)}(\mu)&=&
\tilde{K}^{(r)}(\mu_0)+G^{(r)}(\mu_1)
- \int_{\mu_0}^{\mu_1}
{d \tilde{\mu} \over \tilde{\mu}} \, \lambda_{\tilde{K}}(\tilde{\mu}) \,.
\label{RGE of K and K functions: general}
\end{eqnarray}
We choose the scales
\begin{eqnarray}
\mu_0:=\mu_0(\zeta)= {\zeta P^{-} \over N}  \,, \qquad \mu_1:=\mu_1(\zeta)= e^2 \, \zeta \, P^{-} \,,
\end{eqnarray}
to diminish the initial conditions $\tilde{K}^{(r)}(\mu_0)$ and $G^{(r)}(\mu_1)$.

The evolution kernel $\Gamma$ also contains the diagrams with gluon radiations from the
spectator quark in principle. However, these diagrams contribute at the next-to-leading
logarithmic level, because the NLO effective diagrams with gluon
radiations off the spectator quark do not generate the mixed logarithm $\ln x\,\ln(\zeta^2P^{-2}/k_T^2)$
as indicated by Eqs.~(36) and (37) in \cite{LSW12}. The corresponding soft and hard
functions are expressed as
\begin{eqnarray}
& &K^{\prime}_1 = G_2^{\prime} = K_1, \nonumber\\
& &K^{\prime}_2 \otimes \Phi =  K_2 \otimes \Phi\,,
\nonumber \\
& &G_1^{\prime}={i g^2 C_F \over 2}\int \frac{d^4l}{(2\pi)^4}\frac{ (x \not \! P - \not  l) \not \! u }
{(u \cdot l+i\epsilon) (l^2+i\epsilon) [(x P - l)^2 +i\epsilon]}  \,,
\label{evolution kernel from the  spectator quark}
\end{eqnarray}
which generate
\begin{eqnarray}
{G^{\prime}}^{(b)} &=& G_1^{\prime}-G_2^{\prime}={\alpha_s \, C_F \over 4 \pi}
\left [ {1 \over \epsilon} - \gamma_E + \ln {4 \pi \mu^2 \over \zeta^2(x P^{-})^2 }
- 4 \right ]\,.
\end{eqnarray}
The logarithm $\ln x$ in the soft function
$K^{\prime}_2 \otimes \Phi$ can be extracted by implementing the approximation \cite{LL99}
\begin{eqnarray}
\Phi(x-l^-/P^-,|{\bf k}_T-{\bf l}_T|,\zeta^2,\mu_f) \approx 
\theta(xP^--l^-)\Phi(x,k_T,\zeta^2,\mu_f) \,,
\end{eqnarray}
under which we obtain
\begin{eqnarray}
K_2^{\prime} &=& {i g^2 C_F \over 2} \int \frac{d^4l}{(2\pi)^4}
\frac{ 
\hat u \cdot n  }
{(u \cdot l+i\epsilon)(l^2+i\epsilon) (n_{-} \cdot l+i\epsilon)}
\, \, \theta \left (x P^{-} - l^{-} \right)   \,,\nonumber\\
&=& {\alpha_s \, C_F \over  2 \, \pi} \,
\ln {\zeta \,x P^{-} \over \lambda }  \,.
\label{K2prime: approximation}
\end{eqnarray}
The cancelation of the soft divergences between $K_1^{\prime}\, (=K_1) $
in Eq.~(\ref{K1: result}) and $K_2^{\prime}$ in Eq.~(\ref{K2prime: approximation}) is evident,
whose sum gives
\begin{eqnarray}
{K^{\prime}}^{(b)} &=& K_1^{\prime}+K_2^{\prime}=
-{ \alpha_s \, C_F \over 4 \pi} \left [ {1 \over \epsilon} - \gamma_E
+ \ln \, {4\pi \mu^2   \over \zeta^2 \,  (xP^{-})^2 }    \right ] \,.
\end{eqnarray}

Applying the $\overline{\rm MS}$ scheme to the bare soft and hard functions gives
the renormalized ones
\begin{eqnarray}
{K^{\prime}}^{(r)} &=& -{ \alpha_s \, C_F \over 2 \pi}
\ln \,{ \mu   \over   x\, \zeta \,  P^{-} }   \,,
\\
{G^{\prime}}^{(r)} &=& {\alpha_s \, C_F\over 2 \pi}
\left ( \ln {\mu \over x \, \zeta \, P^{-} }   - 2 \right ) \,.
\end{eqnarray}
Obviously, ${K^{\prime}}^{(r)}$ and ${G^{\prime}}^{(r)}$ are characterized by
the same scale $x \zeta  P^{-}$, implying that a RG treatment is not necessary here, 
and that the sum ${K^{\prime}}^{(r)}+{G^{\prime}}^{(r)}$ produces only a
next-to-leading logarithm as stated above. Hence, this contribution can be absorbed
into the solution of the evolution equation by tuning the
initial rapidity parameter $\zeta$, whose variation within the order-unity
range causes a next-to-leading logarithmic effect. We will take advantage
of the freedom in choosing the bounds of $\zeta$ to achieve the matching between
the resummation formula and the NLO results of the pion transition form factor.
That is, the summation of the above
next-to-leading logarithms can be taken care of by the matching procedure, and
the kernel ${K^{\prime}}^{(r)}+{G^{\prime}}^{(r)}$ will be neglected below.

\subsection{Solution in Mellin and impact-parameter spaces}

Equation~(\ref{evolution equation}) under the Mellin and Fourier transformations
becomes
\begin{eqnarray}
\zeta^2{d \over d \zeta^2} \tilde{\Phi}(N, b, \zeta^2, \mu_f) = \tilde{\Gamma}(N,b, \zeta^2)
\,\,\,  \tilde{\Phi}(N,b, \zeta^2, \mu_f) \,,
\label{evolution equation in Mellin and Fourier space}
\end{eqnarray}
with the evolution kernel
\begin{eqnarray}
\tilde{\Gamma}(N,b, \zeta^2)
&=&\tilde{K}^{(r)}(\mu)+G^{(r)}(\mu)
=  - \int_{\mu_0(\zeta)}^{\mu_1(\zeta)}
\, {d \tilde{\mu} \over \tilde{\mu}} \, \lambda_K(\tilde{\mu}) \,.
\label{evolution kernel in Mellin and Fourier space}
\end{eqnarray}
Solving the differential equation (\ref{evolution equation in Mellin and Fourier space}),
we get
\begin{eqnarray}
\tilde{\Phi}(N, b, \zeta^2, \mu_f)&=&{\rm exp}
\left \{ - \int_{\zeta_0^2}^{\zeta^2} {d \tilde{\zeta}^2 \over\tilde{\zeta}^2 }
\left [ \int_{\mu_0(\tilde{\zeta})}^{\mu_1(\tilde{\zeta})}
{d \tilde{\mu} \over \tilde{\mu}} \lambda_K(\tilde{\mu})
\, \theta \left ( \mu_1(\tilde{\zeta}) - \mu_0(\tilde{\zeta})  \right )
\right  ] \right \}\nonumber\\
& &\times \tilde{\Phi}(N, b, \zeta_0^2, \mu_f) \,,
\end{eqnarray}
which constitutes one of the main  technical results of this paper. The initial rapidity
parameter $\zeta_0$ will be specified later, and the step function in the exponent
will become effective as we perform the inverse Mellin transformation.

Apart from the mixed logarithm, the NLO pion wave function contains
the single logarithm $\ln (\mu_f /Q)$, which can be summed
via the standard RG equation
\begin{eqnarray}
\mu_f\frac{d}{d\mu_f}\tilde{\Phi}(N, b, \zeta^2, \mu_f)
=- \gamma_{\pi}(\mu_f) \, \tilde{\Phi}(N, b, \zeta^2, \mu_f) \,,\label{RGE}
\end{eqnarray}
with the anomalous dimension \cite{LSW11}
\begin{eqnarray}
\gamma_{\pi}(\mu_f)=-{3 \over 2} \, {\alpha_s(\mu_f) \, C_F \over \pi} \,.
\end{eqnarray}
Combining the joint resummation and the solution to Eq.~(\ref{RGE}) leads to
\begin{eqnarray}
\tilde{\Phi}(N, b, \zeta^2, \mu_f)&=&{\rm exp}
\bigg \{ - \int_{\zeta_0^2}^{\zeta^2} {d \tilde{\zeta}^2 \over\tilde{\zeta}^2 }
\, \left [\int_{\mu_0(\tilde{\zeta})}^{\mu_1(\tilde{\zeta})}
\, {d \tilde{\mu} \over \tilde{\mu}} \, \lambda_K(\tilde{\mu})
\, \theta \left ( \mu_1(\tilde{\zeta}) - \mu_0(\tilde{\zeta})  \right )
\right ]
\nonumber \\
&&  \hspace{1.0 cm}  + {3 \over 2} \,
\int_{\mu_i}^{\mu_f} \, {d \tilde{\mu} \over \tilde{\mu}} \,
{\alpha_s(\tilde{\mu}) \, C_F \over  \pi}  \bigg \}
\hspace{0.2 cm}  \tilde{\Phi}(N, b, \zeta_0^2, \mu_i) \,,
\label{resummation of pion wavefunction}
\end{eqnarray}
where $\mu_i$ is the initial scale of the RG evolution.

Note that the physical form factor
\begin{eqnarray}
F(Q^2)=\tilde{\Phi}(N, b, \zeta^2, \mu_f) \otimes \tilde{H} (N, b, \zeta^2, Q^2,  \mu_f) \,,
\label{factorization formula: general}
\end{eqnarray}
is independent of the factorization scheme and the factorization scale $\mu_f$,
where $\tilde{H}$ represents the hard kernel in the Mellin and impact-parameter
spaces. Therefore, we have the evolution equation
\begin{eqnarray}
\zeta^2{d \over d \zeta^2} \, \tilde{H} (N, b, \zeta^2, Q^2,  \mu_f) = - \tilde{\Gamma}(N,b, \zeta^2)
\,\,\,  \tilde{H} (N, b, \zeta^2, Q^2,  \mu_f) \,,
\label{rapidity evolution of hard kernel}
\end{eqnarray}
for the joint resummation, and the RG equation
\begin{eqnarray}
\mu_f\frac{d}{d\mu_f}\tilde{H} (N, b, \zeta^2, Q^2,  \mu_f)
= \gamma_{\pi}(\mu_f) \, \tilde{H} (N, b, \zeta^2, Q^2,  \mu_f) \,.
\label{factorization scale evolution of hard kernel}
\end{eqnarray}
The solution of the above two differential equations gives
the resummation improved hard kernel
\begin{eqnarray}
\tilde{H}(N, b, \zeta^2, Q^2, \mu_f)&=&{\rm exp}
\bigg \{\int_{\zeta^2}^{\zeta_1^2} {d \tilde{\zeta}^2 \over\tilde{\zeta}^2 }
\, \left [\int_{\mu_0(\tilde{\zeta})}^{\mu_1(\tilde{\zeta})}
\, {d \tilde{\mu} \over \tilde{\mu}} \, \lambda_K(\tilde{\mu})
\, \theta \left ( \mu_1(\tilde{\zeta}) - \mu_0(\tilde{\zeta})  \right )
 \right ]
\nonumber \\
&&  \hspace{1.0 cm}  - {3 \over 2} \,
\int^{\mu_f}_{t} \, {d \tilde{\mu} \over \tilde{\mu}} \,
{\alpha_s(\tilde{\mu}) \, C_F \over  \pi}  \bigg \}
\hspace{0.2 cm}  \tilde{H}(N, b, \zeta_1^2,  Q^2, t) \,,
\label{resummation of hard function}
\end{eqnarray}
with the final rapidity parameter $\zeta_1$ and the characteristic hard scale $t$.
We make use of the freedom of choosing the bounds
$\zeta_0^2$ and $\zeta_1^2$ for the joint resummation, such that
the NLO logarithmic enhancements in $\tilde{\Phi}(N, b, \zeta_0^2, \mu_i)$
and $\tilde{H}(N, b, \zeta_1^2, Q^2, t)$, shown in Eqs.~(39) and (40)
of \cite{Nandi:2007qx}, respectively, are eliminated. This requires
\begin{eqnarray}
\zeta_0^2= \left ( {a \, N^{1/4} \over P^- \, b } \right )^{2}\,, \qquad \zeta_1^2= \tilde{a} \,  N^{1/2} \,.
\label{choice of rapidity parameter}
\end{eqnarray}
with the constants
\begin{eqnarray}
a={e^{-1/4} \over 2} \,, \qquad \tilde{a}= (2 \, e)^{-1/2}\,.
\end{eqnarray}

Inserting Eqs. (\ref{resummation of pion wavefunction}) and (\ref{resummation of hard function})
into Eq. (\ref{factorization formula: general}), we derive
\begin{eqnarray}
F(Q^2)&=&{\rm exp}
\bigg \{ - \int_{\zeta_0^2}^{\zeta_1^2} {d \tilde{\zeta}^2 \over\tilde{\zeta}^2 }
\, \left [\int_{\mu_0(\tilde{\zeta})}^{\mu_1(\tilde{\zeta})}
\, {d \tilde{\mu} \over \tilde{\mu}} \, \lambda_K(\tilde{\mu})
\, \theta \left ( \mu_1(\tilde{\zeta}) - \mu_0(\tilde{\zeta})  \right )
\right ]
\nonumber \\
&&  \hspace{1.0 cm}  + {3 \over 2} \,
\int_{\mu_i}^{t} \, {d \tilde{\mu} \over \tilde{\mu}} \,
{\alpha_s(\tilde{\mu}) \, C_F \over  \pi}  \bigg \}
\hspace{0.2 cm}  \tilde{\Phi}(N, b, \zeta_0^2, \mu_i)
\otimes \tilde{H} (N, b, \zeta_1^2, Q^2,  t)\,,  \nonumber \\
&\equiv& \tilde{\Phi}(N, b, \zeta_1^2, t) \otimes \tilde{H} (N, b, \zeta_1^2, Q^2,  t)  \,,
\label{resummation improved form factor: general}
\end{eqnarray}
which recapitulates the joint-resummation improved $k_T$ factorization formula.
The exponential factor in Eq.~(\ref{resummation improved form factor: general})
describes the evolution from the initial condition $\tilde{\Phi}(N, b, \zeta_0^2, \mu_i) $
to the resummation improved wave function $\tilde{\Phi}(N, b, \zeta_1^2, t)$. We have confirmed that
the expansion of the exponential factor up to $O(\alpha_s)$ reproduces the mixed
logarithm and the single logarithm $\ln (1/N)$ in the NLO pion transition form
factor \cite{Nandi:2007qx}. Note that our resummation formalism was established in the
conjugate space, while the calculation in \cite{Nandi:2007qx} was performed in the
momentum space. Hence, the correspondence between $\ln(1/N)$ in the former and $\ln x$
in the latter is not precise, and the matching condition confirmed above in fact suffers
order-unity uncertainty at the next-to-leading-logarithmic level.

At last, we point out that the double logarithm $\ln^2(Q^2/k_T^2)$ in the TMD
wave function and $\ln^2 x$ in the hard kernel were resumed in the conventional
PQCD approach \cite{Nandi:2007qx,TLS}. Besides,
the rapidity parameter $\zeta^2$ was fixed to a specific value for convenience. Compared
to the joint resummation, the double logarithm $\ln^2 x$ in the
TMD wave function has been ignored (see Eq.~(37) in \cite{Nandi:2007qx}
for its existence), and the PQCD approach is not factorization-scheme
independent, strictly speaking. That is, the formalism presented in this work
represents a complete treatment of the logarithmic enhancement in the
pion transition form factor, and the first scheme-independent $k_T$ factorization
formula.

\section{RESUMMATION IMPROVED WAVE FUNCTIONS}
\label{section: resummation improved wavefunctions}

In this section we explore the detailed properties of the resummation improved wave
function $\tilde{\Phi}(N, b, \zeta_1^2, t)$. The factorization theorem for
hard exclusive processes is usually formulated in the momentum-fraction space
(see, however \cite{Bell:2013tfa}).
The inverse Mellin transformation for $\tilde{\Phi}(N, b, \zeta_1^2, t)$ gives
\begin{eqnarray}
\overline{\Phi}(x,b,\zeta_1^2, t)=\int_{c-i\infty}^{c+i\infty} \, \frac{dN}{2\pi i}
\, (1-x)^{-N} \, \tilde{\Phi}(N, b, \zeta_1^2, t),
\label{inverse Mellin transformation}
\end{eqnarray}
where the parameter $c$ is an arbitrary real number larger than the real part of the rightmost
singularity of $\tilde{\Phi}(N, b, \zeta_1^2, t)$  in the complex $N$ plane, and the
Cauchy theorem can be applied to deform the integration contour whenever necessary.
We will not implement the inverse Fourier transformation, so that the joint-resummation
effect can be compared with the Sudakov-resummation effect directly, which is usually
studied in the impact-parameter space.

The parametrization of the TMD pion wave function has been extensively discussed in the literature
(for a recent discussion, see  \cite{Kroll:2010bf}).
For simplicity, factorization of the initial pion wave function in the longitudinal
and transverse momentum spaces
\begin{eqnarray}
\Phi(x,k_T, \zeta_0^2, \mu_i)= \phi(x, \zeta_0^2, \mu_i) \, \, \Sigma(k^2_T) \,,
\label{initial condition: general}
\end{eqnarray}
will be postulated. Keep in mind that the major task of this section is to illustrate
the joint-resummation effect. For definiteness, the transverse momentum
distribution is taken as
\begin{eqnarray}
\Sigma(k^2_T)= 4 \pi \beta^2 \exp(-\beta^2 \, k_T^2) \,,
\end{eqnarray}
where the prefactor is introduced to obey the normalization
\begin{eqnarray}
\int {d^2 k_T \over (2 \, \pi)^2 } \, \Sigma(k^2_T)=1\,,
\end{eqnarray}
and the shape parameter $\beta$ is related to the
root mean square of the transverse momentum via
\begin{eqnarray}
\langle {\bf{k}}_T^2\rangle=\frac{\int^1_0dx\int d^2{\bf{k}}_T
\, {\bf{k}}_T^2 \, |\Phi(x,k_T, \zeta_0^2, \mu_i)|^2}
{\int^1_0 dx \int d^2{\bf{k}}_T \, |\Phi(x,k_T, \zeta_0^2, \mu_i)|^2}
=\frac{1}{2\beta^2}.
\end{eqnarray}
According to \cite{Jakob:1993iw,Lu:2007hr}, the input
$ \langle  {\bf{k}}_T^2  \rangle^{1/2} =350  \, {\rm MeV}$
that fulfills various constraints (including  the $\pi \to \gamma \, \gamma$ decay rate)
leads to $\beta =2.0 \, {\rm GeV^{-1}}$.

The longitudinal momentum distribution $\phi(x, \zeta_0^2, \mu_i)$ is assumed to 
be the same as the light-cone distribution amplitude (LCDA) $\varphi(x, \mu_i)$.
The one-loop evolution equation indicates that the pion LCDA can be
expanded in terms of the Gegenbauer polynomials $ C_n^{3/2}$,
\begin{eqnarray}
\varphi(x, \mu_i) =  6 x (1-x) \sum_{n=0}^{\infty} a_n(\mu_i) \,  C_n^{3/2} (2 x -1) \,,
\end{eqnarray}
where the odd Gegenbauer moments $a_{2n+1}$ vanish due to  symmetry prosperities.
The dependence of $a_{2n}$ on the scale $\mu_i$ is governed by
the well-known Efremov-Radyushkin-Brodsky-Lepage equation \cite{Efremov:1979qk,Lepage:1980fj}.
Along this line, we consider the following three models for the longitudinal momentum distribution
\begin{eqnarray}
\phi^{\rm I}(x, \zeta_0^2, \mu_i)  &=&  6 \, x (1-x)  \,, \nonumber\\
\phi^{\rm II}(x, \zeta_0^2, \mu_i)  &=&  1,\nonumber \\
\phi^{\rm III}(x, \zeta_0^2, \mu_i)  &=&  6 \, x (1-x) \, \left[ 1+ a_2 \, C_2^{3/2}(2 x-1) \right] \,,
\label{initial condition: longitudinal distribution }
\end{eqnarray}
with the Gegenbauer polynomial $C_2^{3/2}(x)=(3/2)(5x^2-1)$.

The first model $\phi^{\rm I}$  corresponds to the pion LCDA in the asymptotic limit.
The flat distribution $\phi^{\rm II}$ was proposed in
\cite{Radyushkin:2009zg,Polyakov:2009je}, where a nonperturbative correction
beyond the operator product expansion was also introduced to explain the scaling violation
indicated by the BaBar data. As there is overwhelming evidence that the pion LCDA
at energy scales accessible in current experiments is
broader than the asymptotic model, we keep the sub-leading Gegenbauer term
in $\phi^{\rm III}$. The contribution from a higher Gegenbauer term
to the pion transition form factor depends on the momentum transfer
squared $Q^2$ and the shape parameter $\beta$ \cite{HWZ13}.
Fitting to the Babar data,  it has been realized that one can at best
determine the second Gegenbauer moment and the shape parameter simultaneously in the framework of the
$k_T$ factorization \cite{Kroll:2010bf}. For this reason, also expecting the quick convergence
of the conformal spin expansion of the pion wave function (see, however, \cite{Agaev:2010aq}),
we will confine the analysis to the second Gegenbauer moment. Our formalism can be extended
to include higher Gegenbauer terms straightforwardly.

\subsection{Resummation with fixed $\alpha_s$}

To make our discussion more transparent, we start from the
inverse Mellin transformation with a frozen coupling constant, and then
generalize it to the case with a running coupling constant.
For a frozen coupling $\alpha_s$, the joint-resummation improved wave function
$\tilde{\Phi}(N, b, \zeta_1^2, t)$ is easily deduced from Eq.~(\ref{resummation of pion wavefunction})
\begin{eqnarray}
\tilde{\Phi}(N, b, \zeta_1^2, t) &=&  \exp \bigg \{ {\alpha_s C_F \over  \pi}
\left [-\ln \left ({\tilde{a} \over a} \, P^{-} b\right )  (\ln N +2) + {3 \over 2}
\ln  \left ({t \over \mu_i} \right) \right ] \bigg \} \,
\nonumber \\
&& \times \tilde{\Phi}(N, b, \zeta_0^2, \mu_i) \,.
\label{pion wavefunction: frozen strong coupling}
\end{eqnarray}
The exponential contains a branch cut on the
negative real $N$ axis and a singularity at $N=0$. The analytical property
of the wave function $\tilde{\Phi}(N, b, \zeta_1^2, t)$ also depends on the initial condition
$\tilde{\Phi}(N, b, \zeta_0^2, \mu_i)$.

\begin{figure}[ht]
\begin{center}
\vspace{1 cm}
\includegraphics[scale=0.4]{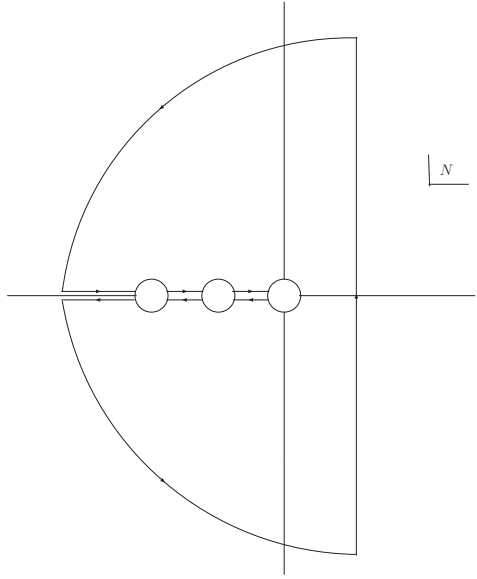}
\caption{Integration contour of the inverse Mellin transformation for the
asymptotic pion wave function. Two  infinitesimal circles  at $N=-1$ and $-2$
and a finite circle at $N=0$  with  radius $r$ are introduced  to ensure that the function
$\tilde{\Phi}(N, b, \zeta_1^2, t)$ is analytical in the region embraced by the contour. }
\label{Integral countour}
\end{center}
\end{figure}

The Mellin and Fourier transformations defined in Eq.~(\ref{Mellin and Fourier transformation})
lead the three models to
\begin{eqnarray}
\label{asy model: initial condition}
\tilde{\Phi}^{\rm I}(N, b, \zeta_0^2, \mu_i)&=& { 6 \over (N+1)(N+2)} \,
\exp\left ( - {b^2 \over 4 \beta^2} \right ) \,,
\\
\tilde{\Phi}^{\rm II}(N, b, \zeta_0^2, \mu_i)&=& { 1 \over N} \,
\exp \left ( - {b^2 \over 4 \beta^2} \right ) \,,
\label{flat model: initial condition}
\\
\tilde{\Phi}^{\rm III}(N, b, \zeta_0^2, \mu_i)&=&
{ 6 \over (N+1)(N+2)}  \left [1 + 6 a_2 \, {(N-1) (N-2) \over (N+3) (N+4)}  \right ]
\, \exp\left ( - {b^2 \over 4 \beta^2} \right ) \,.
\end{eqnarray}
As inserting the asymptotic model
$\tilde{\Phi}^{\rm I}(N, b, \zeta_0^2, \mu_i)$ into
Eq.~(\ref{pion wavefunction: frozen strong coupling}),
$\tilde{\Phi}(N, b, \zeta_1^2, t)$ develops two additional poles at $N=-1$ and $-2$.
According to the Cauchy theorem,  the contour for the inverse Mellin transformation is
deformed  as displayed in Fig.~\ref{Integral countour}, which
(i) runs from minus infinity towards $N=-2$ below the branching cut,
(ii) slides into an infinitesimal semicircle around $N=-2$,
(iii) continues toward $N=-1$ below the branching cut,
(iv) slides into another infinitesimal semicircle around $N=-1$,
(v) runs to $N=-r$ with $0<r<1$,  (vi) revolves around the origin
along a finite circle of radius $r$, and (vii) runs back to minus infinity in a way
that reverses the steps (i)-(v) above the branching cut.

It is trivial to derive the joint-resummation improved pion wave function
\begin{eqnarray}
&& \overline{\Phi}^{\rm I}(x,b,\zeta_1^2, t)
\nonumber \\
\nonumber \\
&& = 6 \exp \left ( - {b^2 \over 4 \beta^2} \right ) \exp \bigg \{ {\alpha_s \, C_F \over  \pi}
\left [- 2 \, \hat{a}   + {3 \over 2} \ln  \left ({t \over \mu_i} \right) \right ] \bigg \}
\nonumber \\
\nonumber \\
&& \times \bigg \{ \sum_{n=1}^{2} \, (-1)^{n-1} \, (1-x)^n \,
\exp \left [ -\alpha_s \, C_F \, \hat{a} \,  \ln n \right ]
\, \cos \left [ \alpha_s \, C_F  \, \hat{a}    \right ]
\nonumber \\
\nonumber \\
&& \hspace{0.4 cm} + \int_{- \pi}^{\pi} \, { d \varphi \over 2 \pi }  \, (1-x)^{- r e^{i \varphi}}
{ r e^{i \varphi}  \over (1+r e^{i \varphi})  (2+r e^{i \varphi}) }  \,
\exp \left [-  {\alpha_s \, C_F \over  \pi} \, \hat{a} \, \ln (r e^{i \varphi})   \right ]
\nonumber \\
\nonumber \\
&& \hspace{0.4 cm} + \int_{\ln r}^{+\infty} {d w \over \pi}  \,  (1-x)^{e^w} {  e^w  \over (1-e^w) (2-e^w) } \,
\exp \left [ -{\alpha_s \, C_F \over  \pi} \,  \hat{a} \, w \right ]
\sin \left [ \alpha_s \, C_F \, \hat{a}  \right ]  \bigg \} \,,
\label{asy model: frozen coupling}
\end{eqnarray}
for the variable
\begin{eqnarray}
\hat{a}= \ln \left ({\tilde{a} \over a}P^{-} b\right )>0 \,.
\end{eqnarray}
The first term in the above expression comes from the contributions of the
$N=-1$ and $-2$ poles,
the second term corresponds to the integration along the circle at $N=0$  with  radius $r$,
and the last term arises from the discontinuity of the integrand along the branching cut.
It can be verified that $\overline{\Phi}^{\rm I}(x,b,\zeta_1^2, t)$ is independent of
the radius $0<r<1$ as it should.

The inverse Mellin transformation is performed along
the same line in the case of the flat model. The initial condition
introduces only a single pole at $N=0$, which is also a branching point of
the exponential in Eq.~(\ref{pion wavefunction: frozen strong coupling}).
The contour is shown in Fig.~\ref{Integral countour}, with
the two infinitesimal circles around $N=-1$ and $=2$ being removed.
This immediately yields
\begin{eqnarray}
&& \overline{\Phi}^{\rm II}(x,b,\zeta_1^2,t)
\nonumber \\
\nonumber \\
&& = \exp \left ( - {b^2 \over 4 \beta^2} \right ) \, \exp \bigg \{ {\alpha_s \, C_F \over  \pi}
\left [- 2  \, \hat{a}   + {3 \over 2} \ln  \left ({t \over \mu_i} \right) \right ] \bigg \}
\nonumber \\
\nonumber \\
&& \hspace{0.5 cm} \times \bigg \{ \int_{- \pi}^{\pi} \, { d \varphi \over 2 \pi }
\, (1-x)^{- r e^{i \varphi}}
\exp \left [-  {\alpha_s \, C_F \over  \pi} \, \hat{a} \,   \ln (r e^{i \varphi})   \right ]
\nonumber \\
\nonumber \\
&& \hspace{1.0 cm} - \int_{\ln r}^{+\infty} \, {d w \over \pi}
\,   (1-x)^{e^w} \, \exp \left [ -{\alpha_s \, C_F \over  \pi}
\, \hat{a} \,  w \right ] \sin \left [ \alpha_s \, C_F \, \hat{a}  \right ] \bigg \} \,.
\label{flat: frozen coupling}
\end{eqnarray}

The resummation improved wave function with the initial condition
$\tilde{\Phi}^{\rm III}(N, b, \zeta_0^2, \mu_i)$ is calculated
similarly, albeit with more involved analytical structures of the integrand;
we need to modify the contour in Fig.~\ref{Integral countour}, so that
two additional poles at $N=-3$ and $-4$ are circumvented.  It
implies that including higher Gegenbauer terms in the initial
condition of the wave function generates a longer sequence
of poles to be avoided in the contour integration. We derive
\begin{eqnarray}
&& \overline{\Phi}^{\rm III}(x,b,\zeta_1^2,t)
\nonumber \\
\nonumber \\
&& = 6 \exp \left ( - {b^2 \over 4 \beta^2} \right ) \exp \bigg \{ {\alpha_s \, C_F \over  \pi}
\left [- 2 \, \hat{a}   + {3 \over 2} \, \ln  \left ({t \over \mu_i} \right) \right ] \bigg \}
\nonumber \\
\nonumber \\
&& \hspace{0.3 cm} \times \bigg \{\sum_{n=1}^{4} \kappa_n \, (1-x)^n   \,
\exp \left [ -\alpha_s \, C_F \, \hat{a} \,  \ln n \right ]
\, \cos \left [ \alpha_s \, C_F  \, \hat{a}   \right ]
\nonumber \\
\nonumber \\
&& \hspace{0.8 cm} + \int_{- \pi}^{\pi} \, { d \varphi \over 2 \pi }  \,
(1-x)^{- r e^{i \varphi}} {r \, e^{i \varphi} \, \, f(a_2, r e^{i \varphi})  \over (1+r e^{i \varphi})  (2+r e^{i \varphi})}
\,
\exp \left [-  {\alpha_s \, C_F \over  \pi}
\, \hat{a} \,  \ln (r e^{i \varphi})   \right ]
\nonumber \\
\nonumber \\
&& \hspace{0.8 cm} + \int_{\ln r}^{+\infty} \, {d w \over \pi}  \,
(1-x)^{e^w} {e^w \, \, f(a_2, -e^w)  \over (1-e^w) (2-e^w) } \, \,
\exp \left [ -{\alpha_s \, C_F \over  \pi}  \,
\hat{a} \,  w \right ] \, \sin \left [ \alpha_s \, C_F  \, \hat{a}  \right ]  \bigg \}  \,,  \nonumber \\
\label{a2 model: frozen coupling}
\end{eqnarray}
where the coefficients $\kappa_n$ and the function $f(x_1,x_2)$  are given by
\begin{eqnarray}
&& \kappa_1=1+ 6 \, a_2\,,  \qquad  \kappa_2=-(1+ 36 \, a_2)\,,
\nonumber \\
&& \kappa_3=60 \, a_2\,, \qquad   \hspace{0.5 cm} \kappa_4=-30 \, a_2\,,
\nonumber \\
&& f(x_1,x_2) = 1 + \, 6 \, x_1 \, {(-1 + x_2) (-2+x_2)  \over (3+x_2) (4+x_2) }  \,.
\end{eqnarray}
Apparently, Eq.~(\ref{a2 model: frozen coupling}) reduces to
Eq.~(\ref{asy model: frozen coupling}), when the second Gegenbauer
moment $a_2$ is set to zero.

A more complicated model for the pion wave function
\begin{eqnarray}
\phi^{\rm IV}(x, \zeta_0^2, \mu_i)  &=&  { \Gamma(2+2 \, \alpha) \over  [\Gamma(1+\alpha)]^2 } \,
{(x \bar x)}^{\alpha} \,,\label{fraction}
\end{eqnarray}
with $0<\alpha<1$, has been advocated in \cite{Cloet:2013tta}. Specifically, the model
with $\alpha=1/2$ was derived in the light-front holographic QCD approach \cite{Brodsky:2013oba}.
The corresponding initial condition
\begin{eqnarray}
\tilde{\Phi}^{\rm IV}(N, b, \zeta_0^2, \mu_i)&=& { \Gamma(2+2 \, \alpha) \over  [\Gamma(1+\alpha)]^2 } \, \,
{ \Gamma(N + \, \alpha)   \over  \Gamma(N + \, 2 \, \alpha +1)} \,
\exp \left ( - {b^2 \over 4 \beta^2} \right ) \,,
\end{eqnarray}
develops infinitely many poles at $N+\alpha=0, \, -1, \, -2, ...$ in the complex $N$ plane.
The study of the joint-resummation effect on this model is similar, and
will not be performed here.

\subsection{Resummation with running  $\alpha_s$}

We are now in a position of computing the joint-resummation improved pion wave function
$\overline{\Phi}(x,b,\zeta_1^2,t)$ with a running coupling $\alpha_s$.
To avoid the Landau singularity in the inverse Mellin transformation,
the parametrization \cite{Solovtsov:1999in}
\begin{eqnarray}
\alpha_s(\mu)={4 \pi \over \beta_0} \left [ {1 \over \ln (\mu^2/\Lambda^2_{\rm QCD})}
- {\Lambda^2_{\rm QCD} \over \mu^2 -\Lambda^2_{\rm QCD}}   \right ],
\label{analytical QCD coupling}
\end{eqnarray}
is adopted at one loop, with the QCD scale $\Lambda_{\rm QCD}$ and
the one-loop QCD $\beta$ function $\beta_0= (11 N_c - 2 N_f)/3$,
$N_c$ and $N_f$ being the numbers of colors and flavors, respectively.
In the above expression
the first term preserves the ultraviolet behavior of the standard QCD coupling,
and the second term cancels the ghost pole at $\mu=\Lambda_{\rm QCD}$.

The substitution of Eq. (\ref{analytical QCD coupling}) into
Eq.~(\ref{resummation of pion wavefunction}) produces
\begin{eqnarray}
\tilde{\Phi}(N, b, \zeta_1^2, t) &=& \exp \left [
{C_F \over \beta_0} ( A_1   + C_1 ) \right ] \,
\tilde{\Phi}(N, b, \zeta_0^2, \mu_i),
\label{pion wf with running coupling: genegral}
\end{eqnarray}
where the functions $A_1$ and $C_1$ are written as
\begin{eqnarray}
A_1&=& \sum_{i=1}^{4} \, (-1)^{n-1} \,\left [  r_i(\ln r_i-1) -  {\rm Li}_2(e^{-r_i}) \right ]\,,
\nonumber  \\
C_1 &=& 3 \left [ \ln {1-  \Lambda^2_{\rm QCD}/\mu_i^2  \over 1-  \Lambda^2_{\rm QCD}/t^2  }
+ \ln { \ln (t^2/\Lambda^2_{\rm QCD})  \over \ln (\mu_i^2/\Lambda^2_{\rm QCD}) }   \right ] \,,
\end{eqnarray}
with the parameters
\begin{eqnarray}
r_{1(3)}&=&{1 \over 2} \, \ln N +\lambda_{1(3)} \,,  \qquad  r_{2(4)}=-{3 \over 2} \, \ln N +\lambda_{2(4)}
\,,\nonumber \\
\lambda_{1(3)} &=& \lambda_{2(4)} +4,   \qquad   \hspace{1.3 cm}
\lambda_2 = 2 \,  \ln { 2 \, a \over  \Lambda_{\rm QCD} \, b} \,,
\qquad   \lambda_4 = 2 \, \ln { 2 \, \tilde{a} P^{-} \over  \Lambda_{\rm QCD}} \,. \nonumber
\end{eqnarray}
The exponential in
Eq.~(\ref{pion wf with running coupling: genegral}) still contains a branching cut
along the negative real $N$ axis, so
the contour in Eq.~(\ref{inverse Mellin transformation})
is deformed in the way exactly the same as in the case with a frozen
coupling.

For the asymptotic pion wave function, we get
\begin{eqnarray}
&& \overline{\Phi}^{\rm I}(x,b,\zeta^2, t)
\nonumber \\
&& =  6 \, \exp \left ( - {b^2 \over 4 \beta^2} \right )
\, \exp \left ( { C_F \over \beta_0}  C_1  \right )
\nonumber \\
\nonumber \\
&& \times \bigg \{ \sum_{n=1}^{2} \, (-1)^{n-1} \, (1-x)^n \,
{\rm exp} \bigg[ F_1(\lambda_1,\lambda_2,\lambda_3,\lambda_4, n) \bigg ]
\, \cos \bigg [ F_2(\lambda_1,\lambda_2,\lambda_3,\lambda_4, n) \bigg ]
\nonumber \\
\nonumber \\
&& \hspace{0.5 cm} + \int_{- \pi}^{\pi} { d \varphi \over 2 \pi } \,  (1-x)^{- r e^{i \varphi}}
\, { r e^{i \varphi}  \over (1+r e^{i \varphi})  (2+r e^{i \varphi}) }
\,\, {\rm exp} \bigg[ F_3(\lambda_1,\lambda_2,\lambda_3,\lambda_4, r e^{i \varphi}) \bigg ]
\nonumber \\
\nonumber \\
&& \hspace{0.5 cm} - \int_{\ln r}^{+\infty} \, {d w \over \pi}  \,  (1-x)^{e^w} {  e^w  \over (1-e^w) (2-e^w) }
\,\, {\rm exp} \bigg[ F_1(\lambda_1,\lambda_2,\lambda_3,\lambda_4,  e^w) \bigg ] \nonumber \\
&&  \hspace{1.0 cm} \times \sin \bigg [ F_2(\lambda_1,\lambda_2,\lambda_3,\lambda_4,  e^w) \bigg ] \bigg \}  \,,
\label{ab1}
\end{eqnarray}
where the explicit expressions of the functions $F_i(\lambda_1,\lambda_2,\lambda_3,\lambda_4, \eta)$
are collected in Appendix \ref{functions F}, and the discontinuity of the
polylogarithm function
\begin{eqnarray}
{\rm Im} \, [{\rm Li}_2 (z \pm i\epsilon)] = \mp \, \pi \, \ln z \,\,  \theta(z-1) \,,
\end{eqnarray}
has been inserted. It has been also verified that the $r$ dependence of
$\overline{\Phi}^{\rm I}(x,b,\zeta_1^2, t)$ cancels between the last two terms
for arbitrary $r$ in the range $0<r<1$.

The same procedure leads to the joint-resummation improved pion wave function
\begin{eqnarray}
\overline{\Phi}^{\rm II}(x,b,\zeta_1^2, t)
&=&  \exp \left ( - {b^2 \over 4 \beta^2} \right )
\exp \left ( { C_F \over \beta_0} C_1  \right )
\nonumber \\
\nonumber \\
&& \hspace{0.2 cm} \times \bigg \{ \int_{- \pi}^{\pi} \, { d \varphi \over 2 \pi } \, (1-x)^{- r e^{i \varphi}}
{\rm exp} \bigg[ F_3(\lambda_1,\lambda_2,\lambda_3,\lambda_4, r e^{i \varphi}) \bigg ]
\nonumber \\
\nonumber \\
&& \hspace{0.7  cm} + \int_{\ln r}^{+\infty} \,  {d w \over \pi}  \,  (1-x)^{e^w} \,
{\rm exp} \bigg[ F_1(\lambda_1,\lambda_2,\lambda_3,\lambda_4,  e^w) \bigg ] \,
\nonumber \\
\nonumber \\
&& \hspace{1.2  cm}  \times \sin \bigg [ F_2(\lambda_1,\lambda_2,\lambda_3,\lambda_4,  e^w) \bigg ] \bigg \} \,,
\label{ab2}
\end{eqnarray}
for the flat model, and
\begin{eqnarray}
&& \overline{\Phi}^{\rm III}(x,b,\zeta_1^2, t)
\nonumber \\
\nonumber \\
&& =  6 \, \exp \left ( - {b^2 \over 4 \beta^2} \right )
\, \exp \left ( { C_F \over \beta_0} C_1 \right )
\nonumber \\
\nonumber \\
&& \hspace{0.5  cm} \times \bigg \{ \sum_{n=1}^{4} \, \kappa_n \, (1-x)^n  \,
{\rm exp} \bigg[ F_1(\lambda_1,\lambda_2,\lambda_3,\lambda_4, n) \bigg ]
\, \cos \bigg [ F_2(\lambda_1,\lambda_2,\lambda_3,\lambda_4, n) \bigg ]
\nonumber \\
\nonumber \\
&& \hspace{1.0  cm} + \int_{- \pi}^{\pi} \, { d \varphi \over 2 \pi }  \, (1-x)^{- r e^{i \varphi}}
\, { r e^{i \varphi}  \over (1+r e^{i \varphi})  (2+r e^{i \varphi}) }
\, {\rm exp} \bigg[ F_3(\lambda_1,\lambda_2,\lambda_3,\lambda_4, r e^{i \varphi}) \bigg ]
\nonumber \\
\nonumber \\
&& \hspace{1.0  cm} \times  \left [1+  6 \, a_2  {(-1 + r e^{i \varphi})
(-2 + r e^{i \varphi})  \over (3 + r e^{i \varphi}) (4 + r e^{i \varphi}) } \right ]
\nonumber \\
\nonumber \\
&&  \hspace{1.0  cm} - \int_{\ln r}^{+\infty} {d w \over \pi}   (1-x)^{e^w} {  e^w  \over (1-e^w) (2-e^w) }
\left [1+  6 \, a_2  {(-1 -e^w) (-2-e^w)  \over (3-e^w) (4-e^w) }   \right ]
\nonumber \\
\nonumber \\
&&  \hspace{1.5 cm} \times {\rm exp} \bigg[ F_1(\lambda_1,\lambda_2,\lambda_3,\lambda_4,  e^w) \bigg ]
\sin \bigg [ F_2(\lambda_1,\lambda_2,\lambda_3,\lambda_4,  e^w) \bigg ] \bigg \}  \,,
\label{ab3}
\end{eqnarray}
for the non-asymptotic model.

\begin{figure}[ht]
\begin{center}
\hspace{-1 cm}
\includegraphics[scale=0.6]{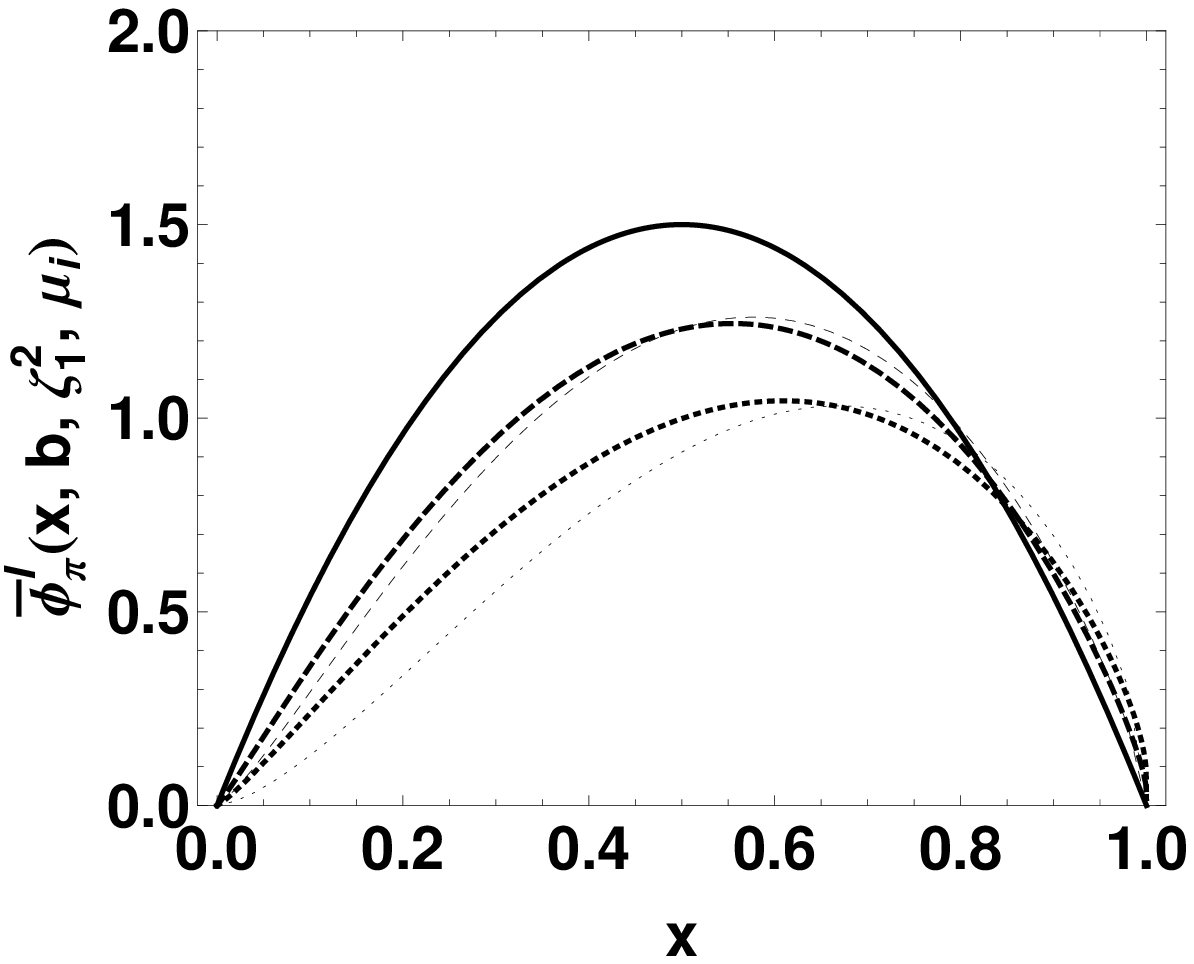}\hspace{0.5 cm}
\includegraphics[scale=0.6]{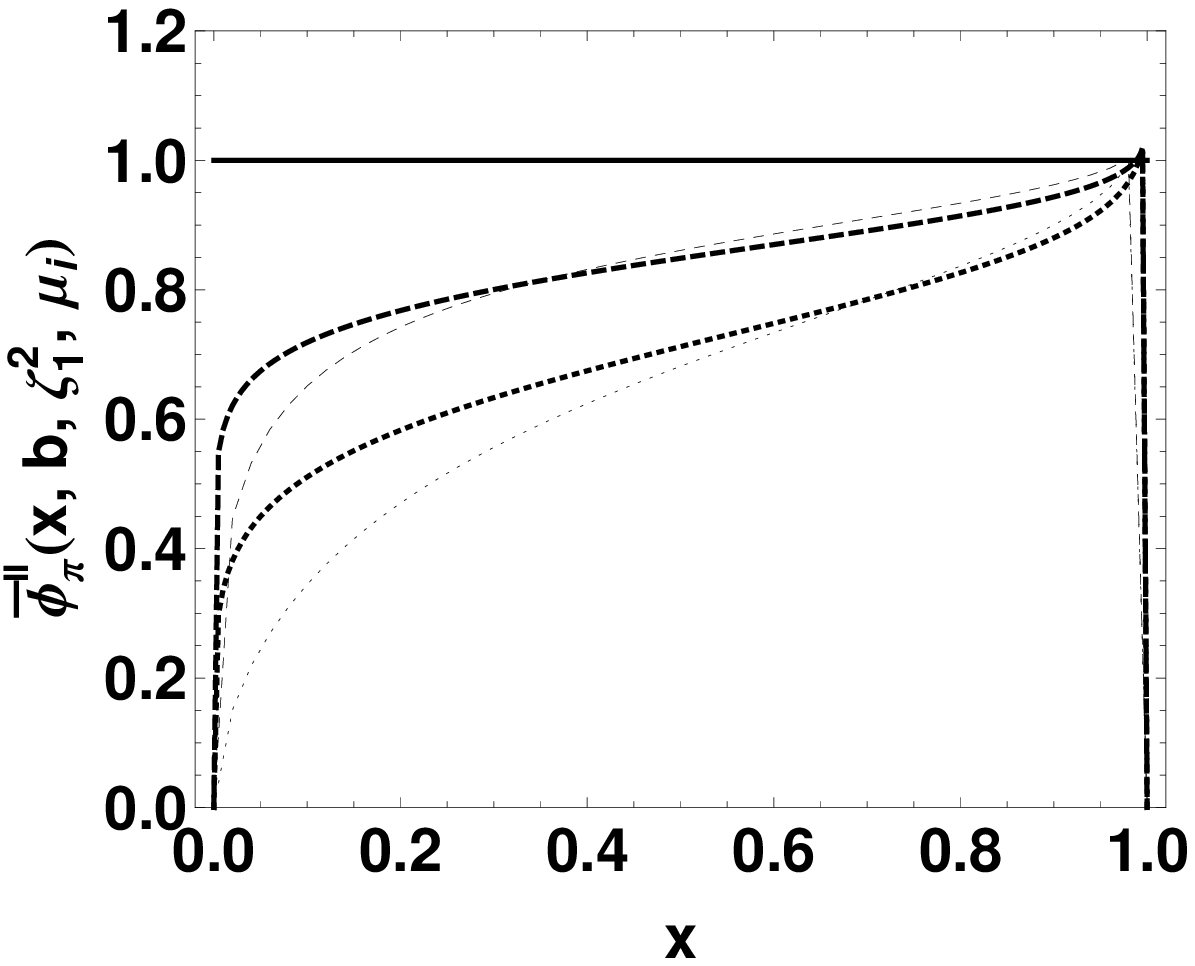} \\
(a)\hspace{7.0cm}(b)\\
\vspace{0.5 cm}
\includegraphics[scale=0.6]{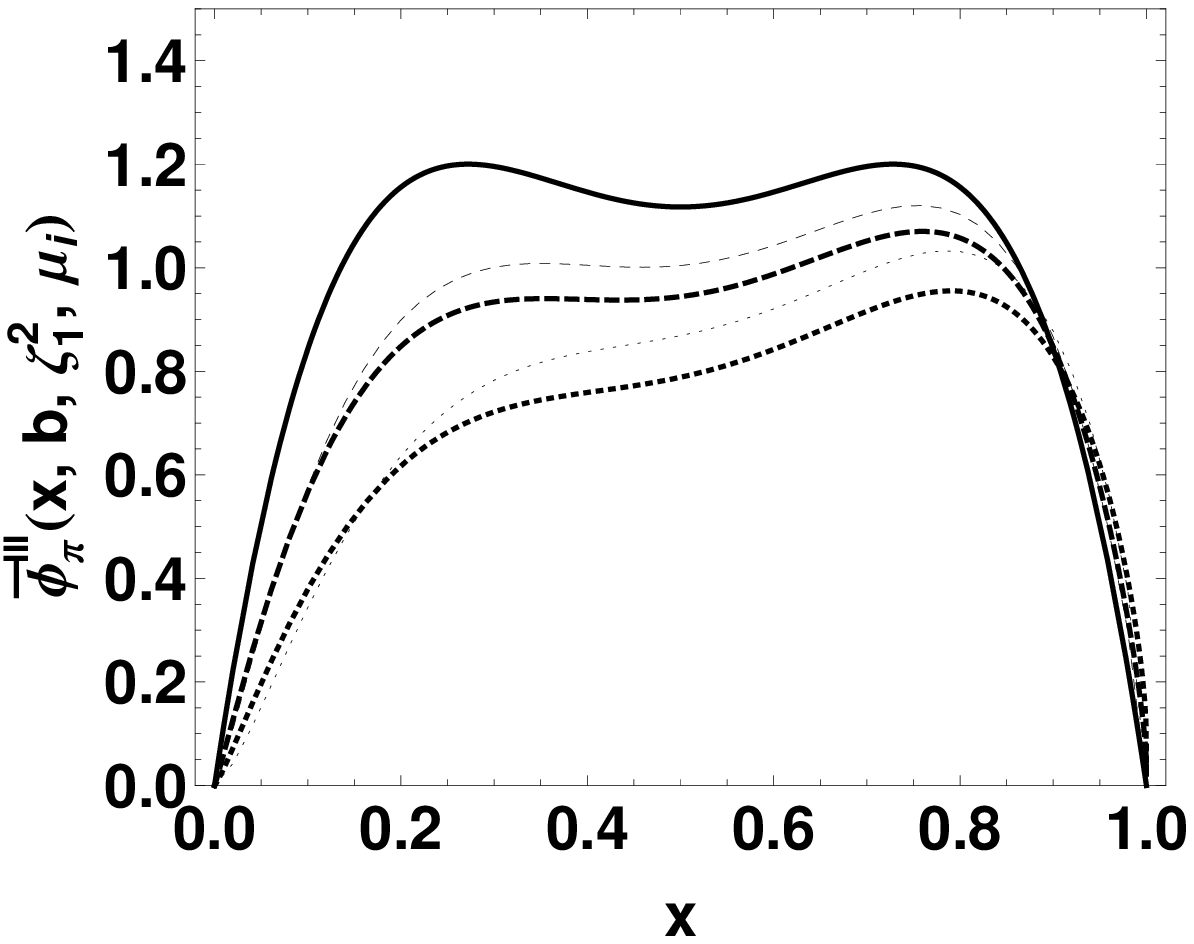}\\
(c)
\caption{Shape of the pion wave function in different models.
(a) the solid, (thin) dashed and  (thin) dotted curves correspond to
the initial condition $\phi^{\rm I}(x, \zeta_0^2, \mu_i)$,
the joint-resummation improved wave function
$\overline{\Phi}^{\rm I}\left(x,{2 \, \tilde{a} \, P^{-}/a},\zeta_1^2,\mu_i\right)$
and $\overline{\Phi}^{\rm I}\left(x,{4 \, \tilde{a} \, P^{-}/a},\zeta_1^2,\mu_i \right)$
for a frozen $\alpha_s=0.3$ (running $\alpha_s$).
(b) the same for the flat pion wave function $\overline{\Phi}^{\rm II}(x,b,\zeta_1^2, \mu_i)$.
(c) the same for the non-asymptotic pion wave function $\overline{\Phi}^{\rm III}(x,b,\zeta_1^2, \mu_i)$
with the second Gegenbauer moment $a_2=0.17$ determined in  \cite{Khodjamirian:2011ub}. }
\label{fig: shape of pion wavefunction}
\end{center}
\end{figure}

The joint-resmmation effect on the pion wave function with a frozen coupling,
for the example set parameters  $\alpha_s=0.3$, $b={2 \, \tilde{a} \, P^{-}/a}$
and  ${4 \, \tilde{a}  \, P^{-}/a} $, and $Q^2=5$ GeV$^2$
is displayed in Fig.~\ref{fig: shape of pion wavefunction}, where the
RG evolution effect is not included.
For all the three considered models, the modification appears as the impact parameter
$b$ is greater than $b_{\rm min}={\tilde{a} / (a \, P^{-})}$, which
is easily understood from the exponentiation of the mixed logarithm
$-\ln \left (\tilde{a} \, P^{-} b/a\right )  (\ln N +2)$
in Eq.~(\ref{pion wavefunction: frozen strong coupling}).
If the rapidity and factorization-scale evolutions are switched off, it is confirmed that
the pion wave function obeys its normalization. A crucial consequence of
the joint resummation, as read from Fig.~\ref{fig: shape of pion wavefunction},
is that the small $x$ region receives stronger suppression compared to the
moderate $x$ region as expected, while the large $x$ region almost remains intact.
Moreover, the suppression strengthens with the transverse separation $b$ between the
valence quarks at a given longitudinal momentum fraction. Therefore, the
joint-resummation effect does not allow a significant contribution from soft gluon
exchanges. This well known Sudakov mechanism, first formulated
in QED, improves the applicability of PQCD to hard exclusive
processes.

Turning to the case with a running coupling, we adopt the parameter $\Lambda_{\rm QCD}=
250$ MeV and the flavor number $N_f=6$. As observed from
Fig.~\ref{fig: shape of pion wavefunction}, the resummation improved pion wave function
takes on a behavior rather similar to that for a frozen coupling.
A minor difference is that the small $x$  region is even more suppressed in the
former case, which further boosts our confidence on
the applicability of PQCD to exclusive processes at
moderate momentum transfer.

\begin{figure}[ht]
\begin{center}
\includegraphics[scale=0.6]{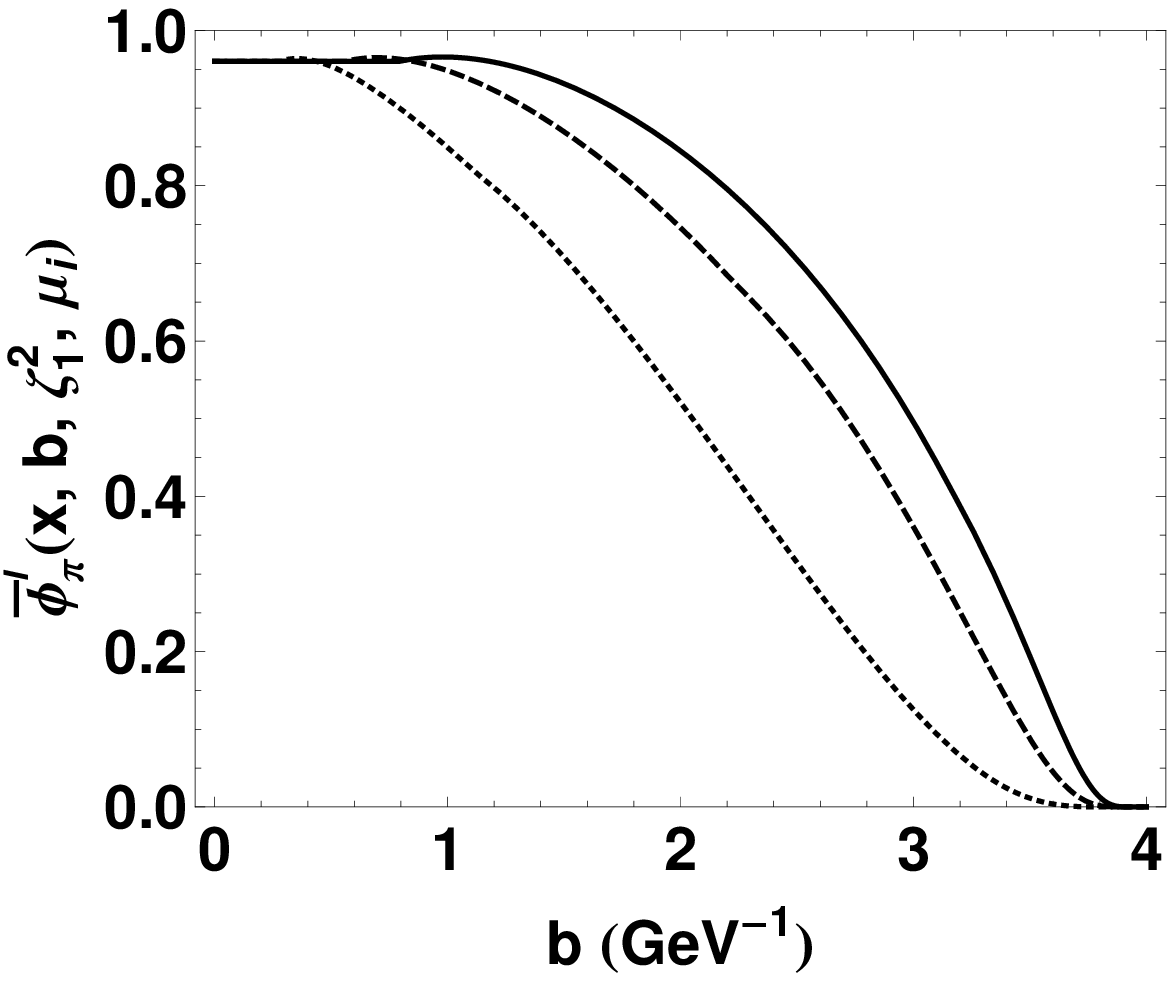}\hspace{0.5 cm}
\includegraphics[scale=0.6]{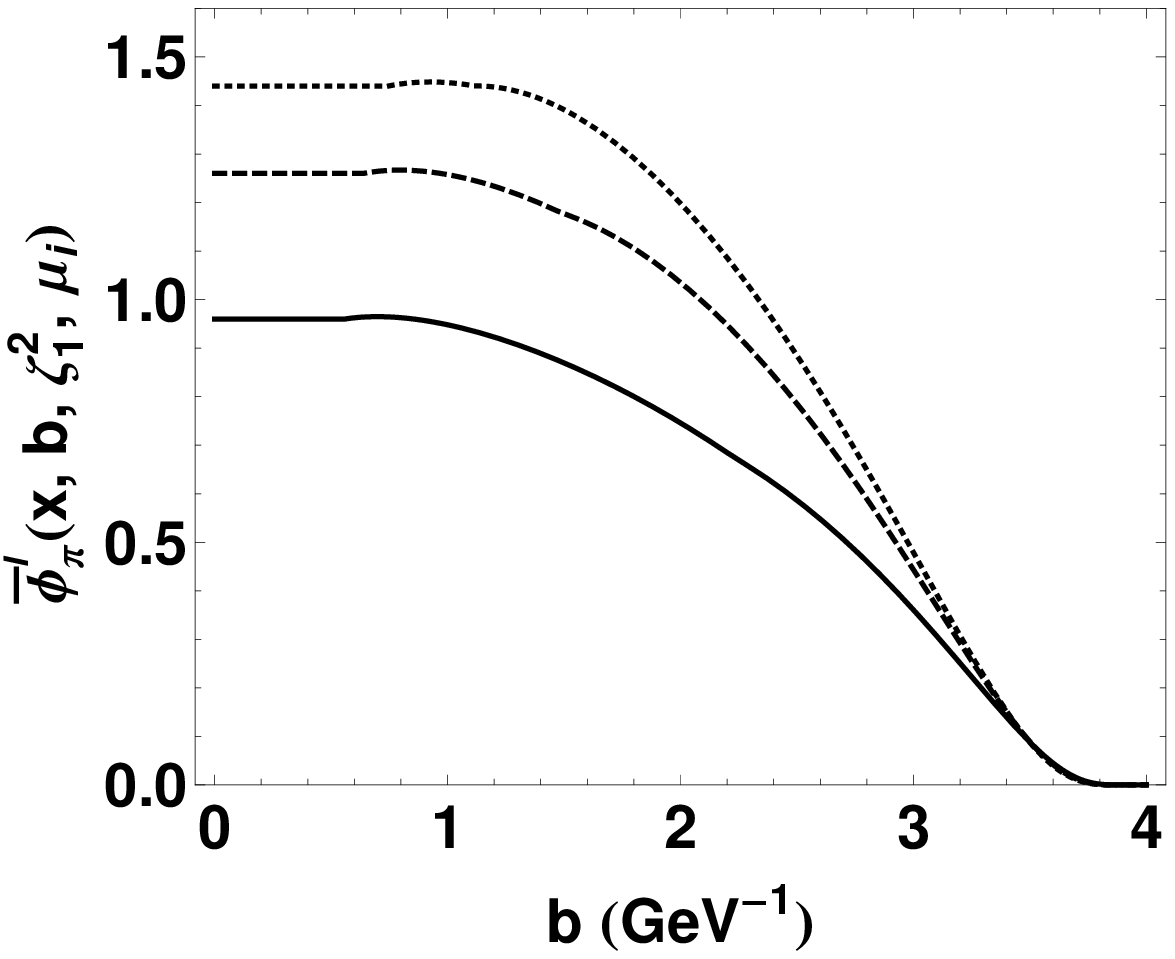}
\\(a)\hspace{7.0cm}(b)\\
\vspace{1 cm}
\includegraphics[scale=0.6]{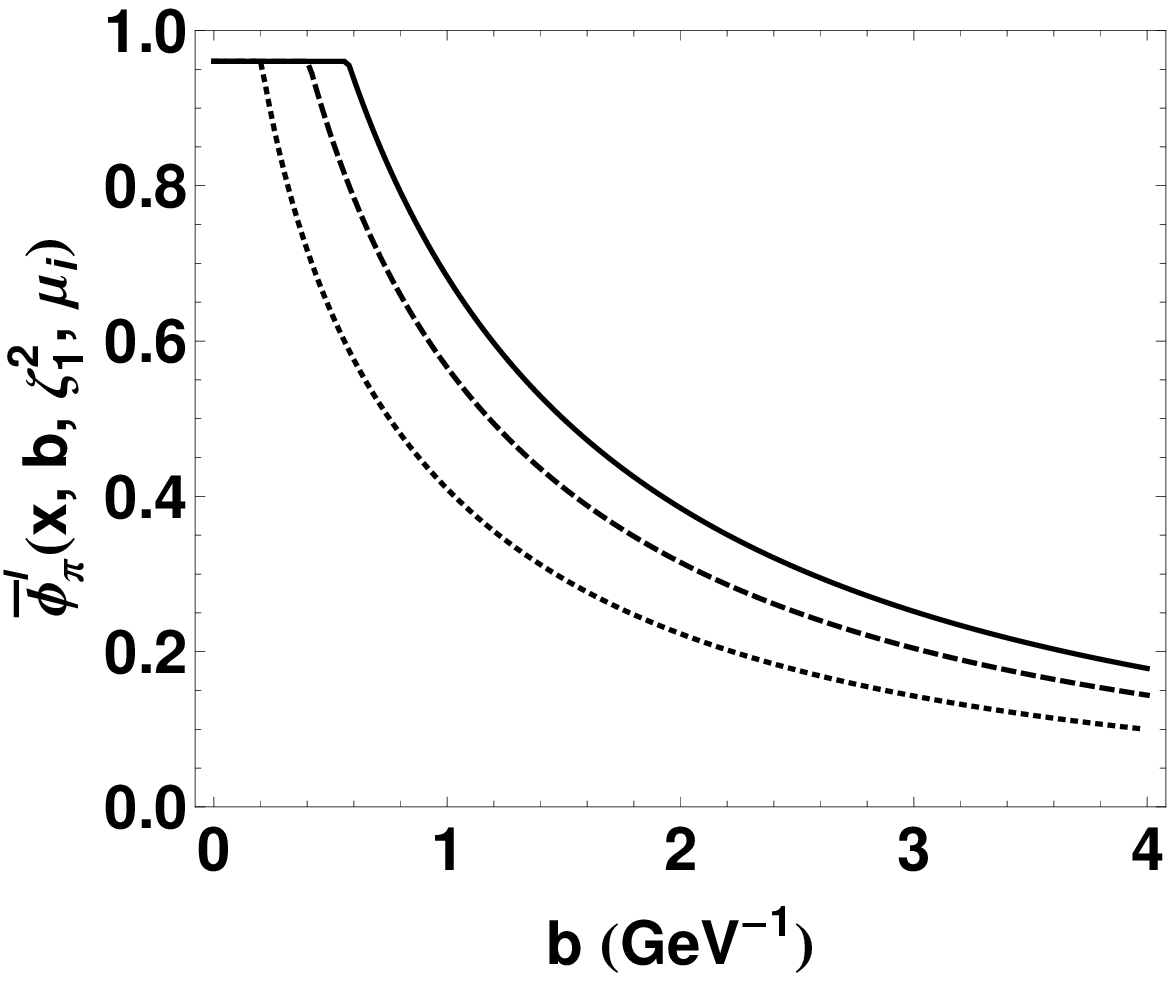}\hspace{0.5 cm}
\includegraphics[scale=0.6]{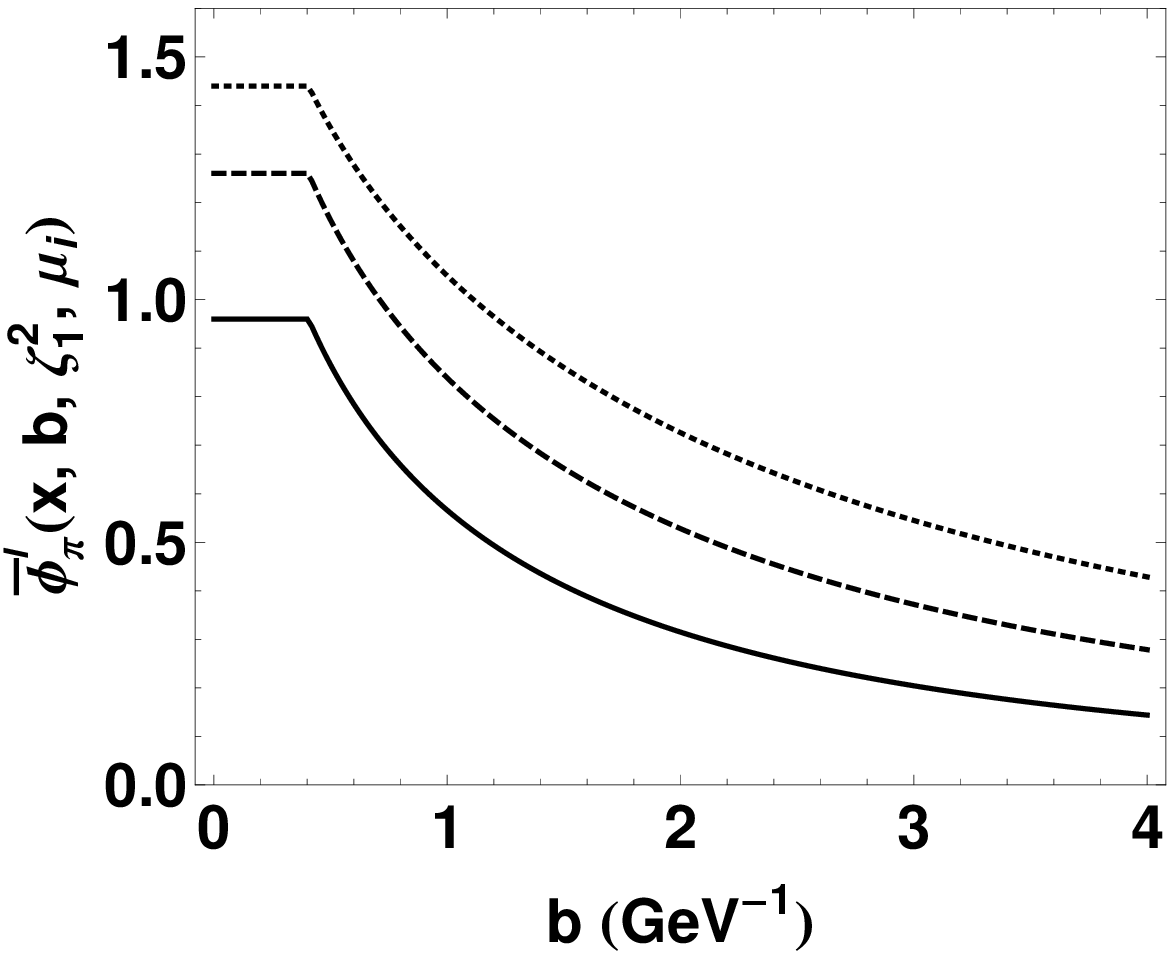}\\
(c)\hspace{7.0cm}(d)
\caption{Distinction between the asymptotic pion wave functions
including the Sudakov resummation and including the joint resummation.
The solid, dashed, and dotted curves correspond
to the Sudakov-resummation improved pion wave function (a) at
$Q^2=5$ ${\rm GeV}^2$, 10 ${\rm GeV}^2$,  and 40 ${\rm GeV}^2$
for the momentum fraction  $x=0.2$,
and (b) at $Q^2=10$ ${\rm GeV}^2$ for $x=0.2$, $0.3$, and $0.4$.
The same for the joint-resummation improved pion wave functions in (c) and (d).}
\label{fig: compare two resummations}
\end{center}
\end{figure}

Before closing this section, we highlight the distinction between the pion wave functions
including the Sudakov resummation and including the joint resummation.  For simplicity,
we confine ourselves to the asymptotic model, because the other two models exhibit a
similar $b$ dependence. As indicated in Fig.~\ref{fig: compare two resummations},
both resummation formalisms lead to suppression on the wave function in the large
$b$ region, which becomes more significant as the momentum transfer $Q$ increases.
This observation fulfills the concept that an energetic pion is a compact
hadronic bound state. A striking feature is that the
joint-resummation improved wave function concentrates on the
small $b$ region more than the Sudakov-resummation improved one, which
falls off smoothly in the intermediate $b$ region. The concentration
on the small $b$ region is attributed to the exponentiation of the mixed
logarithm $-\ln \left (\tilde{a} \, P^{-} b/a\right )\ln N $, where the
suppression with $b$ is magnified by the large coefficient $\ln N$. However, the
Sudakov-resummation improved wave function vanishes quickly as $b\to 1/\Lambda_{\rm QCD}$,
since the running coupling $\alpha_s(1/b)$ hits the Landau pole. In the present derivation
the Landau pole has been avoided as shown in Eq.~(\ref{analytical QCD coupling}), such
that the joint resummation does not diminish the wave function as
$b\to 1/\Lambda_{\rm QCD}$. We emphasize that this distinction, due to the
different treatments of the Landau-pole contribution, is not physically crucial.
It is found that the large $b$ region is more suppressed with the growing of
$x$ in Figs.~\ref{fig: compare two resummations}(b),
but not in Fig.~\ref{fig: compare two resummations}(d): the small-$x$ approximation
has been adopted in the joint resummation, so its effect is insensitive to the
variation of $x$. The phenomenological consequences
on the pion transition form factor from the two resummations will be elaborated
in the next section.

\section{PION TRANSITION FORM FACTOR}
\label{section: pion transition form factor}

The pion transition form factor $F(Q^2)$ involved in the $\gamma^{\ast} \pi^{0} \to \gamma$ process
is defined by the following matrix element
\begin{eqnarray}
\langle \gamma(P^{\prime},\epsilon^*) | j_{\mu}^{em}(q)  | \pi^{0}(P)\rangle
= i \, g_{em}^2  \, \varepsilon_{\mu \nu \alpha \beta}  \epsilon^{*\nu}   P^{\alpha} P^{\prime \beta} F(Q^2) \,,
\end{eqnarray}
where $j_{\mu}^{em}(q)$ is an electromagnetic current, $q=P^\prime-P$ is the momentum transfer,
and $\epsilon$ denotes the polarization of the outgoing photon.
The form factor $F(Q^2)$ ($Q^2=-q^2$) was written, in the collinear factorization, as \cite{Lepage:1980fj}
\begin{eqnarray}
F(Q^2)= {\sqrt{2} \, f_{\pi} \over 3} \int_0^1 dx { \varphi(x,t)  \over  x Q^2}
\left [ 1 + H^{(1)}(x, Q^2, t) \right ] \,,
\end{eqnarray}
with the NLO hard kernel \cite{delAguila:1981nk,Braaten:1982yp,Kadantseva:1985kb}
\begin{eqnarray}
H^{(1)}(x, Q^2, t) ={\alpha_s(t) C_F \over 2 \pi}
\left [ - \left ( \ln x +{3 \over 2} \right ) \ln {t^2 \over Q^2}
+ {1 \over 2} \ln^2 x  -{ x \ln x \over 2 (1-x)} -{9 \over 2} \right ]  \,.
\end{eqnarray}
It is seen that $F(Q^2)$  scales as $1/Q^2$ from the power counting
of the hard kernel, and is determined by the inverse moment of the pion LCDA at LO.

\subsection{$k_T$ factorization formula}

To suppress the end-point contribution (soft gluon exchanges) from the small
$x$ region in the collinear factorization,
the $k_T$ factorization has been developed for hard exclusive processes,
and continually refined by including the resummations of important logarithms and power corrections
as stated in the Introduction. This more sophisticated factorization theorem can be
derived diagrammatically \cite{Nagashima:2002ia} by applying the eikonal approximation
to collinear particles and the Ward identity to the diagram summation in the leading infrared
regions. For the rapidity  parameter $\zeta^2=2$, the $k_T$ factorization formula
at leading power of $1/Q^2$
under the conventional resummations was given by  \cite{Nandi:2007qx,LM09}
\begin{eqnarray}
F(Q^2) &=& {\sqrt{2} \, f_{\pi} \over 3} \int_0^1 dx  \int_0^{\infty} b \, db \, \overline{\Phi}(x, b,t)
\,\, e^{-S(x,b,Q,t)} \,\, S_t(x,Q)   \,  \nonumber \\
&& \times K_0(\sqrt{x} Q \, b)   \left [1-  {\alpha_s(t) C_F \over 4 \pi}
\left ( 3 \ln {t^2 \, b \over 2 \sqrt{x} Q} + \gamma_E + 2 \ln x + 3 -{\pi^2 \over 3}   \right ) \right ]  \,.
\label{conventional PQCD}
\end{eqnarray}
The Sudakov factor $S(x,b,Q,t)$ sums the double logarithm
$\ln^2 ({k_T^2 / Q^2})$ and the single logarithm $\ln ({t^2 / Q^2})$ through
the RG equation. The threshold factor from the resummation of $\ln^2 x$ has been parameterized as
\begin{eqnarray}
S_t(x,Q) &=&{2^{1+c(Q^2)} \, \Gamma ({3 \over 2}+c(Q^2))  \over   \sqrt{\pi} \, \Gamma(1+c(Q^2)) }
\, \left [ x (1-x) \right ]^{c(Q^2)}   \,,  \label{threshold}
\end{eqnarray}
for convenience, in which the power $c(Q^2)$ was determined to be
\begin{eqnarray}
c(Q^2) &=& 0.04 Q^2-0.51 Q + 1.87 \,,
\end{eqnarray}
by fitting to the exact threshold resummation formula in the Mellin space.
It was then observed that the nontrivial $Q^2$ dependence of $c(Q^2)$ is important
for accommodating both low and high $Q^2$ data from BaBar.
Note that the self interactions of the Wilson links have been included into the
NLO hard kernel in Eq.~(\ref{conventional PQCD}),
such that the coefficient of the first term in the brackets has been changed from ``$1$'' to ``$3$'',
compared to Eq. (40) in \cite{Nandi:2007qx}.
As argued in \cite{Li:2008hu}, the additional contribution from
these self interactions can be canceled by the soft subtraction in an alternative definition
for the TMD pion wave function, as the involved gauge parameter is tuned appropriately.
In this work  we have adopted the definition of the TMD pion wave function with off light-cone Wilson links.

To minimize the factorization-scheme dependence of the pion transition form factor, the
resummation of the mixed logarithms in both the pion wave function
and the hard kernel has been performed in the previous section. The large logarithms
in the initial conditions of the pion wave function and the hard kernel were eliminated
by choosing the bounds $\zeta_0^2$ and $\zeta_1^2$
in Eqs. (\ref{resummation of pion wavefunction}) and (\ref{resummation of hard function}),
leading to
\begin{eqnarray}
H^{(1)}(x,k_T, \zeta_1^2,  Q^2, t)&=& -{\alpha_s(t) C_F \over 4 \pi}
\bigg ( 3 \ln {t^2 \over  x Q^2+k_T^2} + \ln 2 +2  \bigg ) \,.
\end{eqnarray}
We then arrive at the joint-resummation improved factorization formula for the pion transition form factor
\begin{eqnarray}
F(Q^2) &=& {\sqrt{2} \, f_{\pi} \over 3} \int_0^1 dx  \int_0^{\infty} b \, db \, \overline{\Phi}(x, b, \zeta_1^2, t)
K_0(\sqrt{x} Q \, b)  \,  \nonumber \\
&& \times\left [1-  {\alpha_s(t) C_F \over 4 \pi}
\left ( 3 \ln { t^2 \, b \over 2 \sqrt{x} Q} + \ln 2 + 2   \right ) \right ]   \,,
\label{joint resummation PQCD}
\end{eqnarray}
with $\overline{\Phi}(x, b, \zeta_1^2, t)$ coming from Eqs.~(\ref{ab1}),
(\ref{ab2}), and (\ref{ab3}).

\subsection{Numerical analysis}

\begin{figure}[H]
\begin{center}
\includegraphics[scale=0.48]{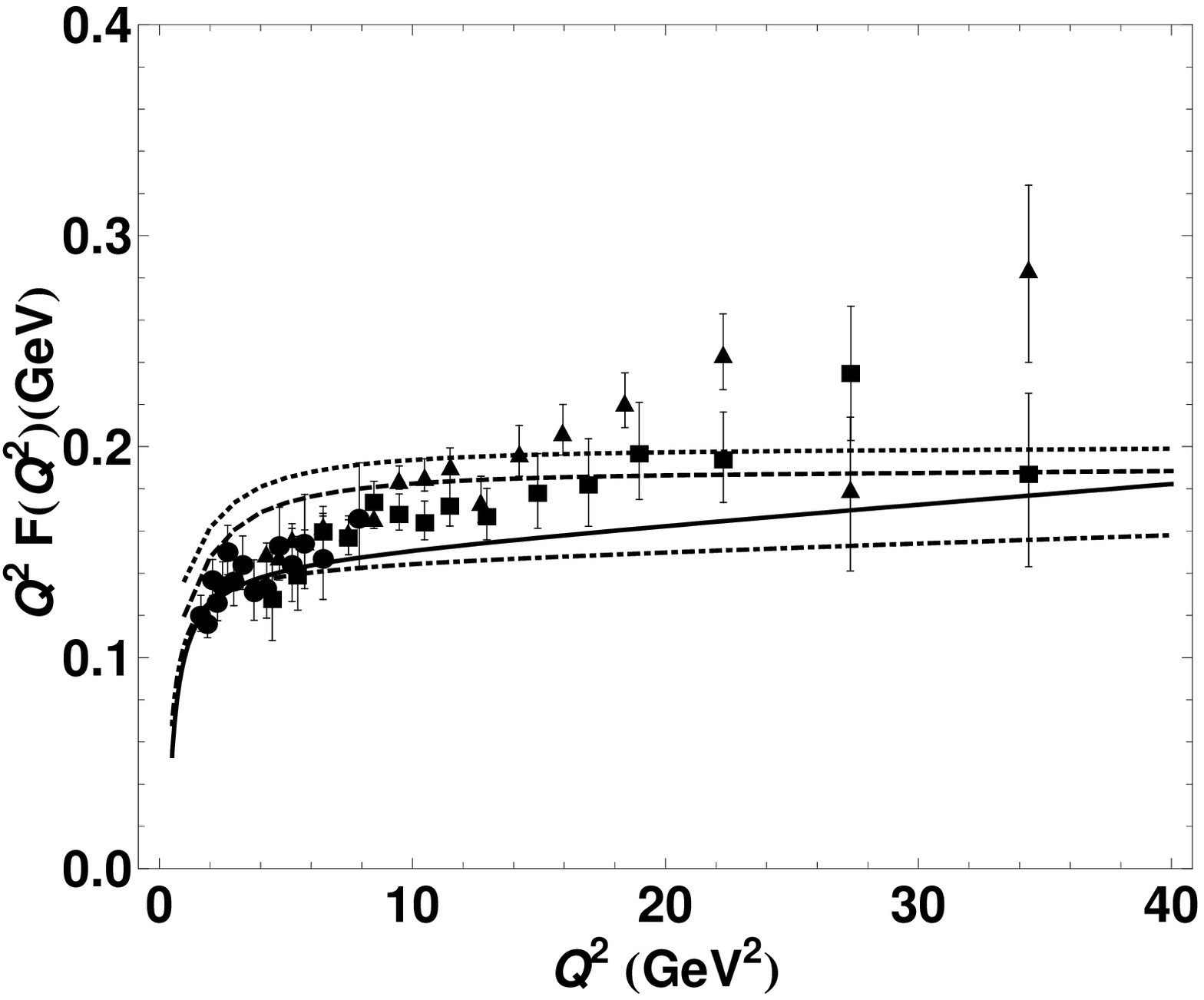}\\
(a)\\
\includegraphics[scale=0.48]{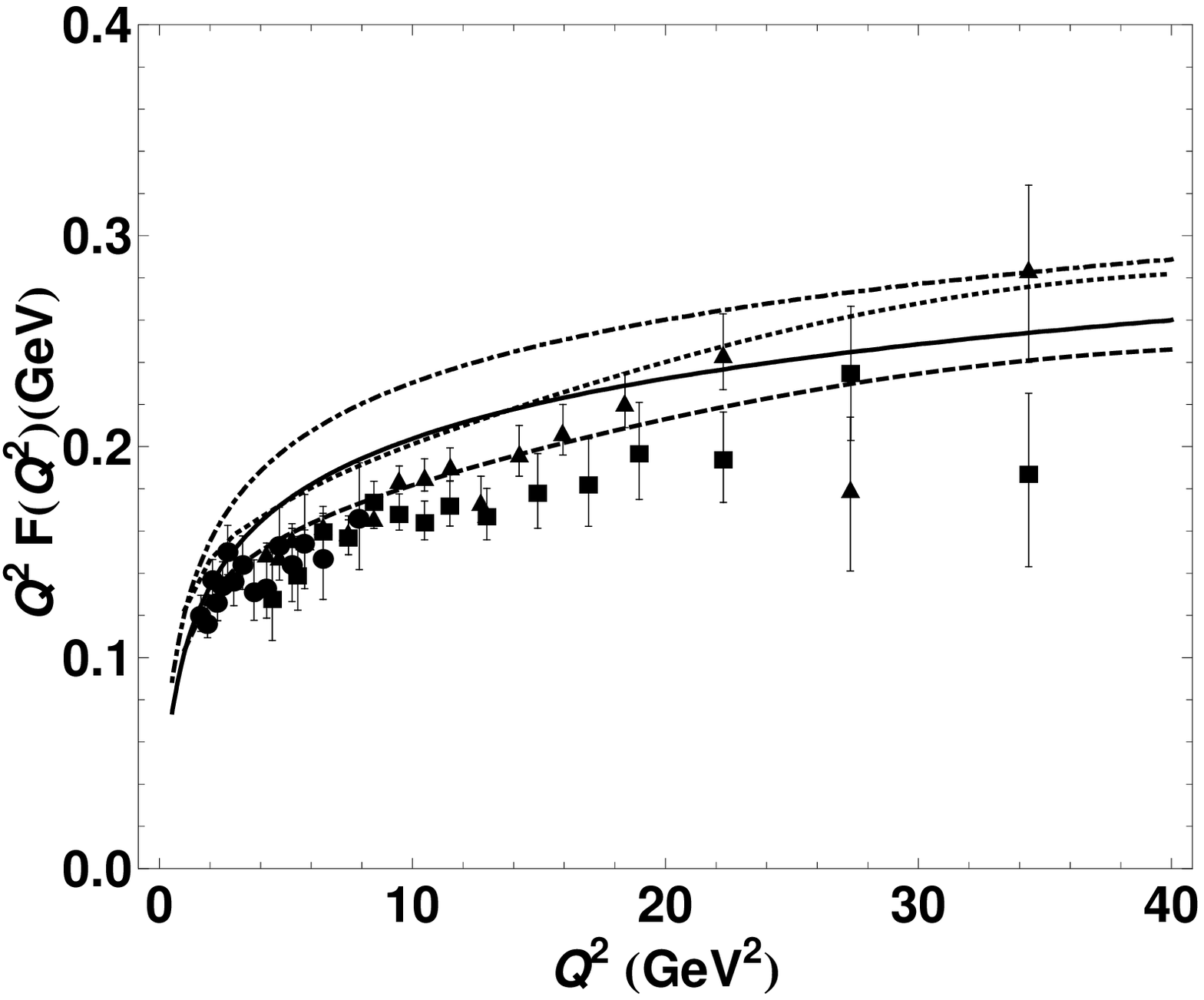}\\
(b)\\
\includegraphics[scale=0.48]{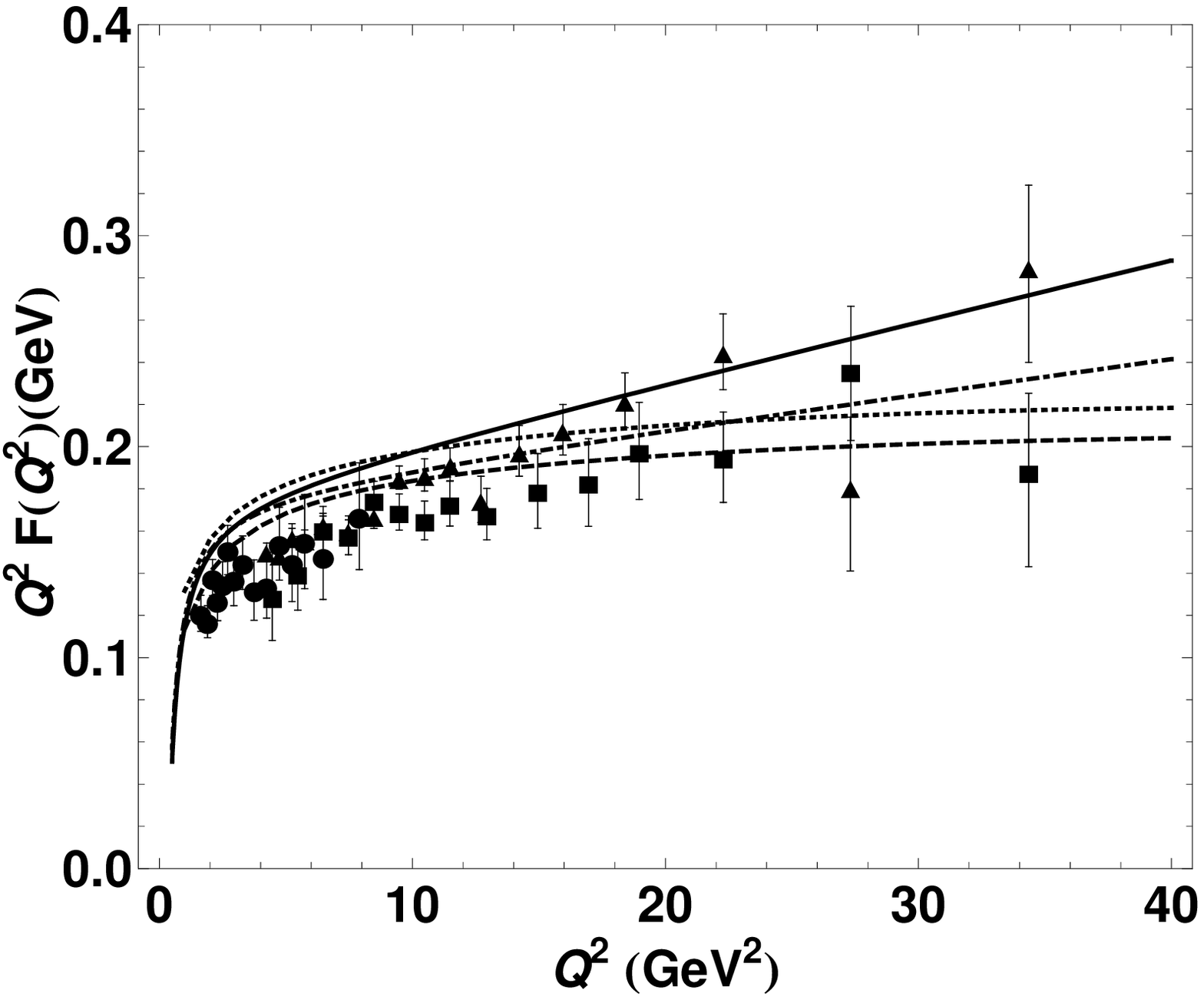}\\
(c)
\caption{Pion transition form factor calculated from (a) the asymptotic model,
(b) the flat model, and (c) the non-asymptotic model.
The dashed and dotted  (dot-dashed and solid)  curves indicate the LO and NLO
predictions from the conventional resummations (joint resummation).
The  experimental data are from CLEO \cite{Gronberg:1997fj} (dots),
BaBar \cite{Aubert:2009mc} (triangles),
and Belle \cite{Uehara:2012ag} (squares).   }
\label{fig: pion form factor}
\end{center}
\end{figure}

The first issue in the numerical analysis concerns the choice of the hard characteristic
scale $t$. One choice would be $t^2=\sqrt{x} Q/b$ that removes the remaining logarithm 
in Eq.~(\ref{joint resummation PQCD}). Another choice characterizing the
typical quantum fluctuation of hard scattering is $t={\rm max} (\sqrt{x} Q, 1/b)$ as widely adopted
in the PQCD approach \cite{LS}. In both scenarios the hard scale runs into the nonperturbative
region at small $x$ and large $b$, but take different values. We have confirmed that
the two scenarios do not generate practical difference in our
formalism for the pion transition form factor.
It implies that the joint resummation has suppressed the contribution from
the nonperturbative region effectively.
Below we will adopt the second scenario as the default choice.

Another important issue is the determination  of the Gegenbauer moment $a_2$.
QCDSR calculations of moments of the pion wave function  can be traced back to 1980s, pioneered
by Chernyak and  Zhitnitsky \cite{Chernyak:1981zz}, where a rather high value $a_2(\mu) \sim 0.58$
at the scale $\mu^2 \in [1, \, 2] \, {\rm GeV}^2$ was derived. This estimate was improved gradually
by including NLO QCD corrections and refining the ``internal" parameters  of the QCDSR approach.
The most recent update gave $a_2(1 \, {\rm GeV} ) = 0.15 \pm 0.03$ \cite{Bakulev:1998pf}.
Following the strategy of LCSR, we will not use the Gegenbauer coefficient $a_2$
computed from QCDSR directly. Instead, we employ the value \cite{Khodjamirian:2011ub}
\begin{eqnarray}
a_2 (1 \, {\rm GeV})= 0.17 \pm 0.08 \,,
\end{eqnarray}
extracted from matching the LCSR evaluation of the pion electromagnetic form factor,
which includes NLO twist-2 corrections
and higher power terms up to twist 6, to the experimental data from the Jefferson Lab Collaboration.

We start our numerical analysis with the input of the asymptotic pion wave function,
choosing the initial scale $\mu_i=1/b$ for the RG evolution in
Eq.~(\ref{resummation of pion wavefunction}) \cite{LS}.
As observed from Fig.~\ref{fig: pion form factor}(a),
the predicted $Q^2F(Q^2)$ with the conventional resummations at both LO and NLO
levels saturates rapidly as $Q^2 > 5 \, {\rm GeV^2}$, and the NLO QCD correction
enhances the form factor by $(6 - 14) \%$. It is clear that the asymptotic
pion distribution generally accommodates the Belle data except the first two bins. However,
it cannot describe the CLEO and BaBar data in both small and large $Q^2$ region.
Note that the incompatibility between the BaBar and Belle data on the pion transition
form factor has been elaborated quantitatively in \cite{Stefanis2}.
Besides, the impact at low energy of
the BaBar and Belle high-energy data was analyzed by means of
the Pad\'{e} approximation inspired parametrization in \cite{Masjuan:2012wy}.
The joint-resummation effect decreases the LO and NLO predictions in the conventional
approach by $(11 - 16) \%$ and $(8 -27) \%$, respectively.
Such decrease can be understood via the stronger reduction
at small $x$ from the joint resummation as shown in
Figs.~\ref{fig: compare two resummations}(b)
and \ref{fig: compare two resummations}(d), which is the dominant region owing to the
hard kernel $K_0(\sqrt{x} Q \, b)$. It is found that the saturation behavior of $Q^2F(Q^2)$
changes slightly at NLO in the joint-resummation improved $k_T$ factorization:
the NLO correction brings about $6 \%$ suppression ($15 \%$ enhancement)
to the LO result in the small (large) $Q^2$ region. The above decrease of the NLO
form factor at small $Q^2$ is explained as follows. The contribution to the
pion transition form factor under the joint resummation mainly comes from
the small $b$ region as indicated in Fig.~\ref{fig: compare two resummations}.
We then have $t^2b \sim 1/b > \sqrt{x}Q$ at small $Q$, for which the logarithm of
the NLO hard kernel in Eq.~(\ref{joint resummation PQCD}) flips sign.
The failure of describing the experimental data suggests that the pion wave function might be broader.
This observation is in agreement with
the particular feature of the pion as a Nambu-Goldstone boson of dynamical chiral symmetry breaking
and with the recent lattice simulations \cite{Cloet:2013tta}.

The computed pion transition form factor $Q^2F(Q^2)$ with the flat pion wave function is exhibited in
Fig.~\ref{fig: pion form factor}(b). It is seen that the form factor grows steadily with $Q^2$
at LO and NLO in both resummation formalisms.
This is easily realized from the scaling $Q^2 F(Q^2) \sim \ln (Q^2/k_T^2)$ implied by the
tree-level $k_T$ factorization formula with the flat pion wave function \cite{LM09}.
The LO curve from the conventional approach reasonably describes the scaling violation at
large $Q^2$ observed by BaBar and the low $Q^2$ data from CLEO and Belle.
The NLO correction increases the form factor by approximately $(15 - 18) \%$, such that
the agreement with the data deteriorates a bit.
Compared to the conventional approach, the predictions from the joint resummation
brings about $17 \%$ enhancement and $(8-16) \%$ suppression at the LO and NLO levels,
respectively. The enhancement at LO, opposite to what was observed in
Fig.~\ref{fig: pion form factor}(a) for the asymptotic model, arises from the
weaker suppression than the parameterized threshold factor in Eq.~(\ref{threshold})
at small $x$. The NLO correction becomes destructive under the joint resummation
in the whole range of $Q^2$, decreasing the LO result by $(10-15) \%$.
This behavior, different from that in the case of the asymptotic model,
is also traced back to the logarithmic term in Eq.~(\ref{joint resummation PQCD}):
$\ln (t^2 b/(\sqrt{x} Q))$ remains positive in the small $x$ region, which is probed
more by the flat pion wave function. It is interesting to notice that the NLO curves from
the conventional resummations and from the joint resummation turn out to be similar.

The input of the third model of the pion wave function with a nonvanishing second Gegenbauer
moment $a_2$ leads to $Q^2F(Q^2)$ displayed in Fig.~\ref{fig: pion form factor}(c).
The form factor under the conventional resummations behaves in a way similar to that in
Fig.~\ref{fig: pion form factor}(a): $Q^2F(Q^2)$ saturates as $Q^2 > 10 \, {\rm GeV^2}$, and the
magnitude is larger; namely, it goes between the BaBar and Belle data. These features
are attributed to the broader pion wave function, which enhances the small-$x$ contribution.
Compared to the form factor in the conventional approach, it shows $(4 - 18) \%$ enhancement
at LO and $(13 - 32) \%$ correction at NLO under the joint resummation. With the strong
suppression at the end point, the major contribution does not come from small $x$, but
from moderate $x$, say, $0.1 < x <0.2$. In this range the pion wave function takes values
$\overline{\Phi}^{\rm I}<\overline{\Phi}^{\rm II}<\overline{\Phi}^{\rm III}$ as revealed
in Fig.~\ref{fig: shape of pion wavefunction}. It explains why the curves for $Q^2F(Q^2)$
under the joint resummation ascend fastest with $Q^2$ in Fig.~\ref{fig: pion form factor}(c),
a result not expected from \cite{Brodsky:2011yv,Melikhov,Stefanis,Brodsky:2011xx}.

\begin{figure}[H]
\begin{center}
\includegraphics[scale=0.48]{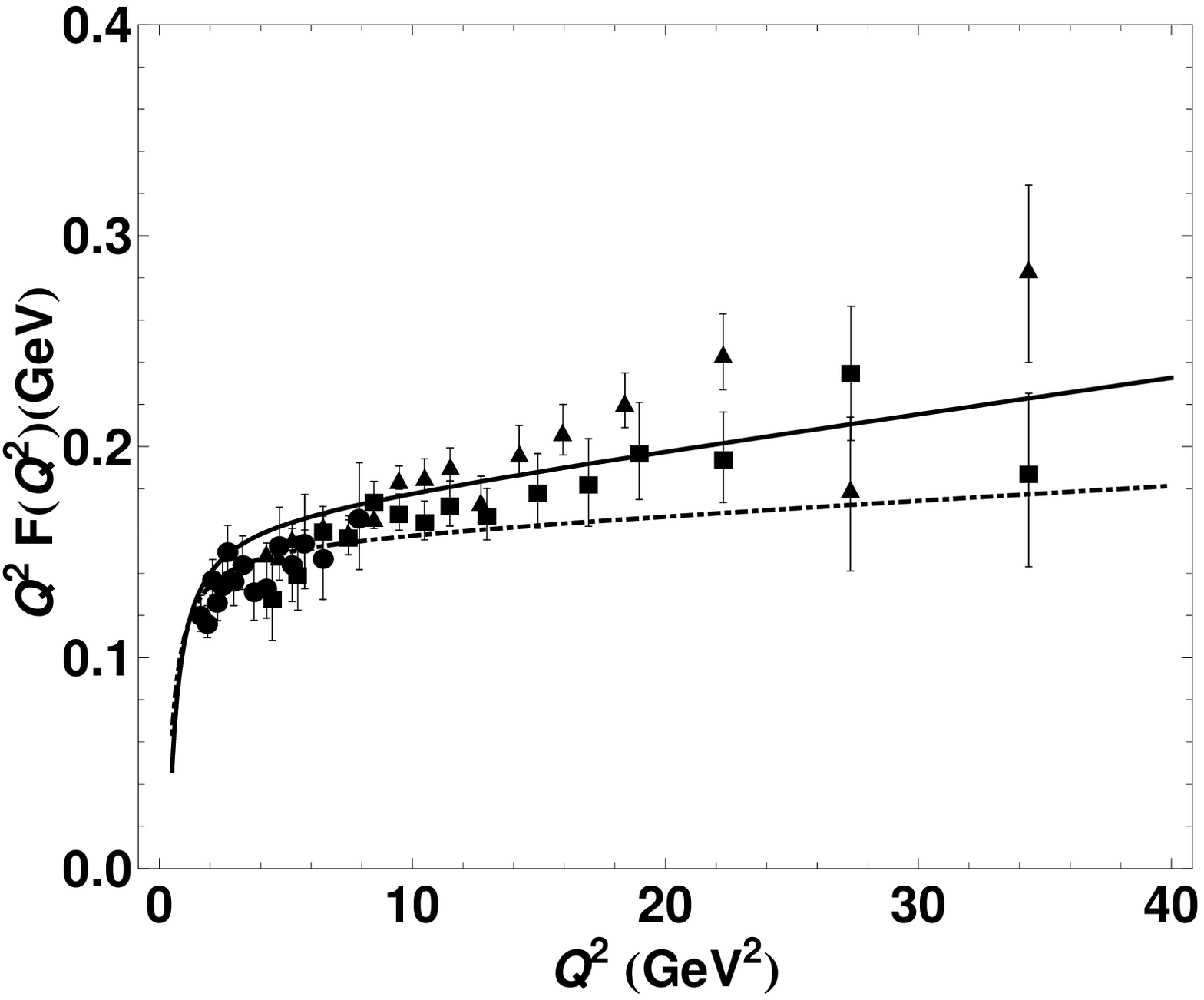}\\
\caption{LO (dot-dashed) and NLO (solid) pion transition form factor
calculated from the non-asymptotic model with the second Gegenbauer moment $a_2=0.05$
under the joint resummation.
}
\label{fig:a2}
\end{center}
\end{figure}

The above reasoning implies that a pion wave function with two humps, such as the
Chernyak-Zhitnitsky model \cite{Chernyak:1981zz} or the Bakulev-Mikhailov-Stefanis model
\cite{BMS}, which further lift the values in $0.1<x<0.2$, will overshoot the BaBar data
in our formalism. To improve the description of the data, instead,
a smoother pion wave function, lying between the
asymptotic and flat models in the range $0.1<x<0.2$,
serves the purpose as hinted by Figs.~\ref{fig: pion form factor}(a) and
\ref{fig: pion form factor}(b). The examples include a non-asymptotic model with
a smaller Gegenbauer moment $a_2$ and a model in Eq.~(\ref{fraction}) with a
fractional power $\alpha<1$. We present the LO and NLO results for $Q^2F(Q^2)$
under the joint resummation and the input of the pion wave function with
$a_2=0.05$  \footnote{Such a value of the second Gegenbauer moment still lies in
a very conservative bound $0 \leq a_2 (1 \, {\rm GeV}) \leq 0.3$  proposed in
Eq. (9) of \cite{Ball:2005tb}. Moreover, $a_2$ in the TMD pion
wave function needs not to to be the same as in the pion LCDA.
}  in Fig.~\ref{fig:a2}.
Fairly speaking, this model wave function describes
reasonably well the CLEO, BaBar, and Belle data in the whole range of $Q^2$.
In summary, the significance of the NLO contribution and the saturation behavior
of the pion transition form factor are quite different under the joint
resummation and the conventional resummations. Therefore, it is important to
have appropriate treatment of QCD logarithmic corrections to a process, before its
data are used to extract a hadron wave function.
It is also crucial to clarify the high $Q^2$ data of the pion transition form factor on
the experimental side, in order to acquire
better understanding of the hadron structure and stringent scrutinization of perturbation theory.

\begin{figure}[H]
\begin{center}
\includegraphics[scale=0.48]{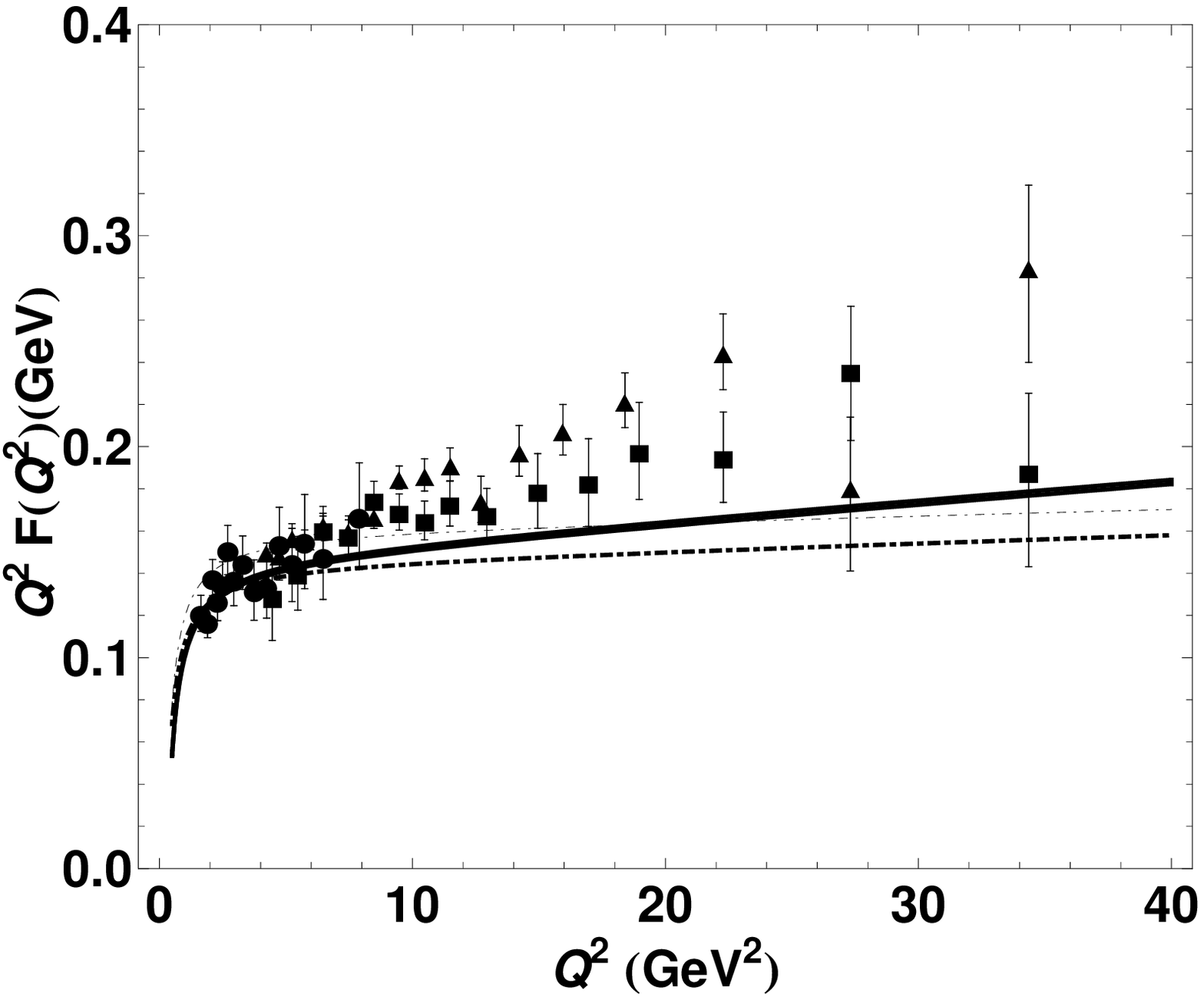}\\
(a)\\
\includegraphics[scale=0.48]{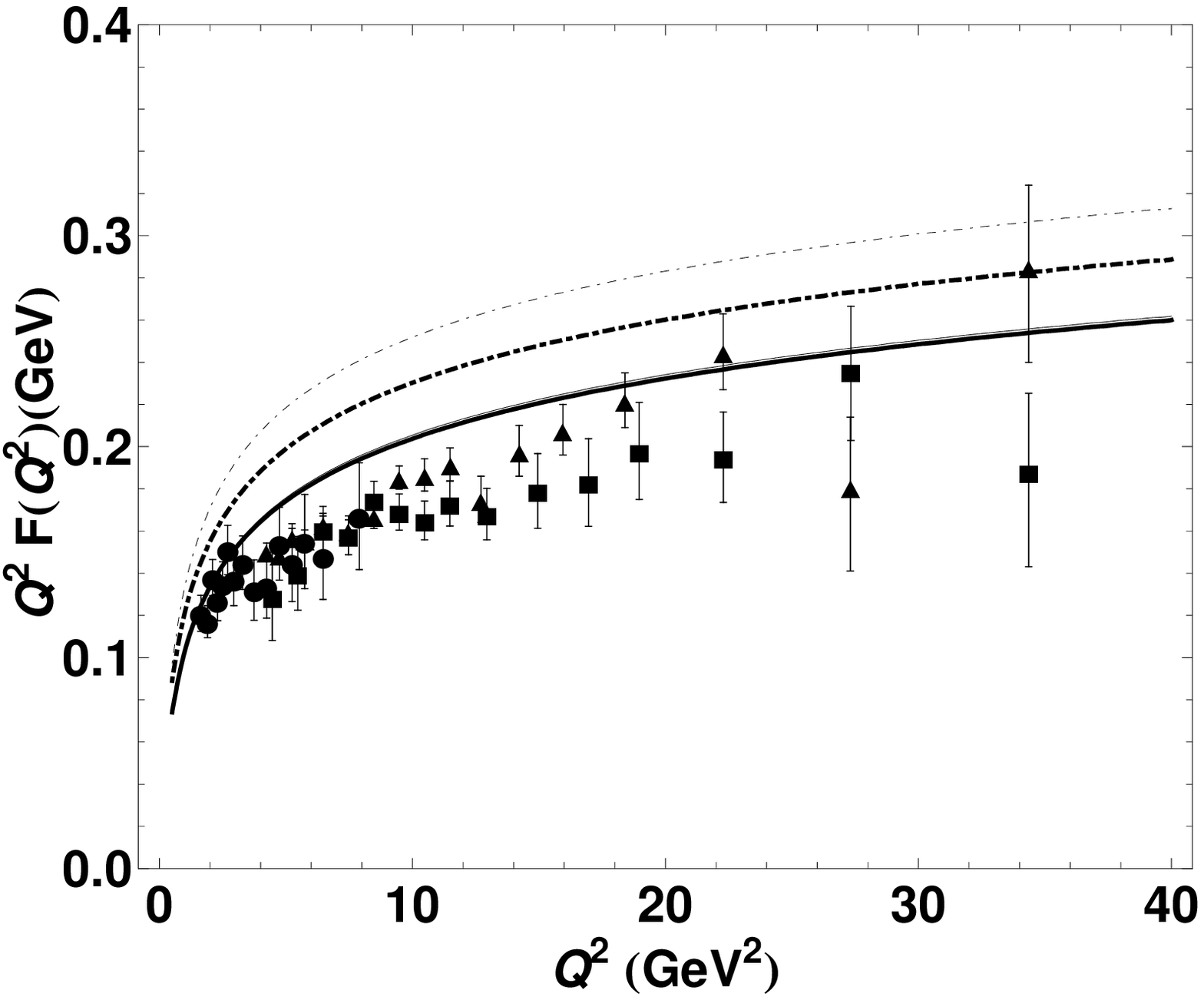}\\
(b)\\
\includegraphics[scale=0.48]{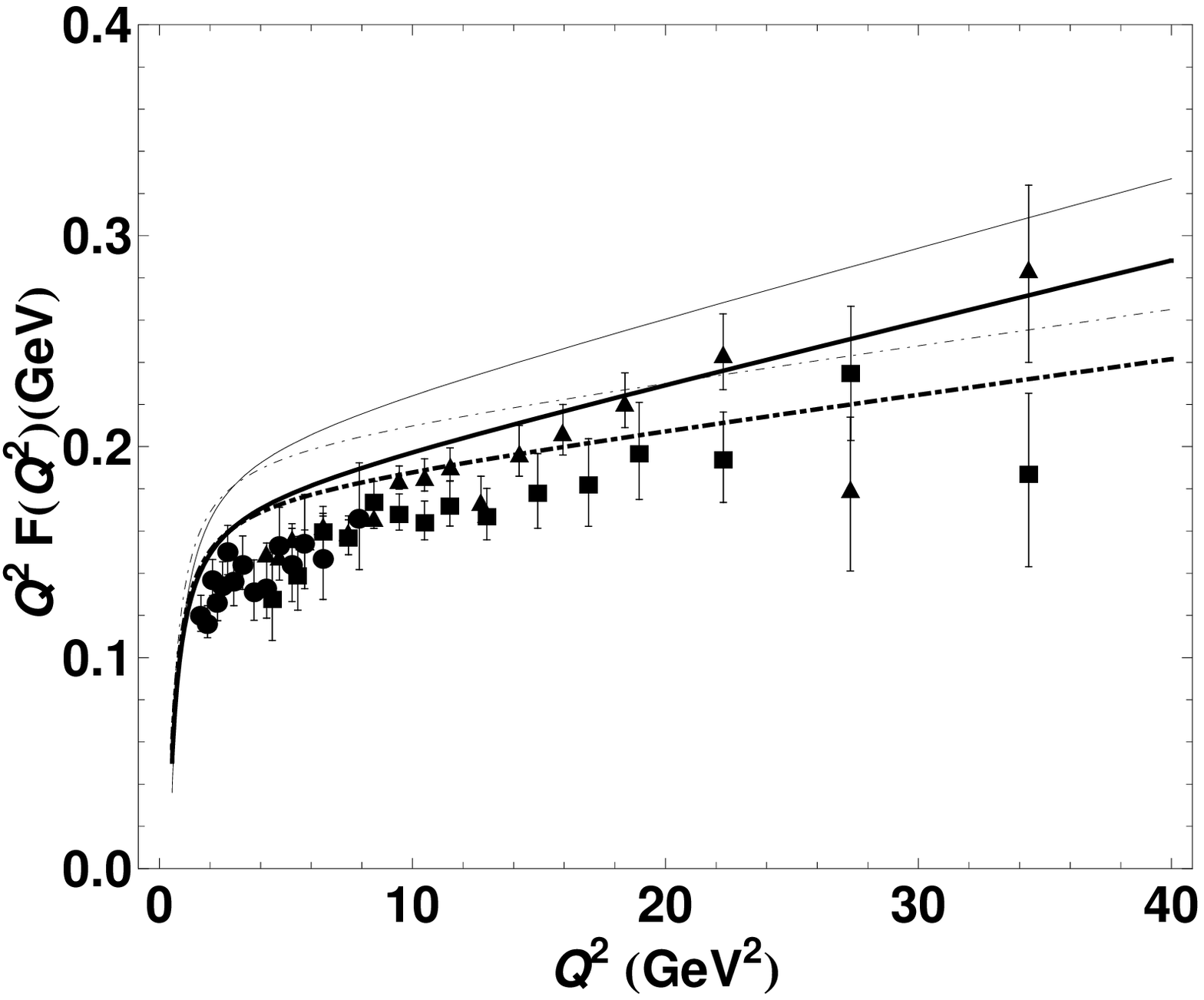}\\
(c)
\caption{Hard scale induced uncertainties of the pion transition form factor
for (a) the asymptotic model, (b) the flat model, and (c) the non-asymptotic model.
The (thin) dot-dashed and solid
curves correspond to the LO and NLO predictions, respectively, under the joint resummation
with $t={\rm max} (\sqrt{x} Q, 1/b)$ ($t={\rm max} (2 \, \sqrt{x} Q, 1/b)$). }
\label{fig: pion form factor uncertainties}
\end{center}
\end{figure}

To illustrate theoretical uncertainties, we vary the default choice of the hard scale
into $t={\rm max} (2 \,\sqrt{x} Q, 1/b)$.
As shown in Fig.~\ref{fig: pion form factor uncertainties}, the scale variation increases the pion transition form
factor in the large $Q^2$ region by approximately $8 \%$  and $1 \%$ at  LO and NLO, respectively,
for the asymptotic pion wave function.
On the other hand, tuning the hard scale magnifies the QCD correction, as large as $7 \%$,
to the pion transition form factor in the flat model.  This is certainly not unexpected, taking into account
the highlighted role of the single logarithm  $\ln (t^2 \, b /(\sqrt{x} Q))$ in the hard kernel.
Similar observation also holds for the non-asymptotic model with a finite $a_2$, albeit with the
NLO correction being enhanced to $13 \%$ at NLO. The range of the above numerical results are
basically consistent what was obtained in \cite{Stefanis2}.

\section{CONCLUSION AND DISCUSSION}
\label{section: conclusion}

Applying the resummation technique with off-light-cone Wilson lines, we have constructed
an evolution equation to resum the mixed logarithm $\ln x\,\ln(\zeta_P^2/k_T^2)$ in the TMD pion wave function.
The joint-resummation improved pion wave function modifies both the longitudinal and transverse momentum
distributions. As a consequence, the moderate $x$ and small $b$ regions are more highlighted compared
to the case with the conventional threshold and $k_T$ resummations.
We stress that the joint resummation, organizing all the important logarithms in the pion
wave function and in the hard kernel, is a treatment more appropriate and complete than the
conventional resummations. In particular, Eq.~(\ref{joint resummation PQCD})
derived in this work represents the first scheme-independent
$k_T$ factorization formula for the pion transition form factor in the presence of
the light-cone singularity.

We have examined the significance of the NLO contribution and the saturation behavior
of the pion transition form factor at high energy under the joint resummation.
Differences from those under the conventional resummations were noticed, indicating
that QCD logarithmic corrections to a process must be handled
appropriately, before its data are used to extract a hadron wave function.
Our predictions for the pion transition form factor have been confronted  with the
measurements from CLEO, BaBar and Belle by testing three models for
the pion wave function. The comparison shows that a smooth pion wave function
is favored over a pion wave function with two humps in our formalism. It turns out
that a non-asymptotic pion wave function with a small second Gegenbauer moment
$a_2=0.05$ describes reasonably well the CLEO, BaBar, and Belle data in the whole
range of $Q^2$. Resolving the discrepancy between the BaBar and Belle measurements will definitely
improve our understanding towards the hadronic structure of a pion.


Our scheme-independent formalism can be extended to the $k_T$ factorization
of more complicated exclusive processes. We will demonstrate this extension taking
the pion electromagnetic form factor as an example. The first step is to
verify that the choice of $\zeta_0^2$, with $N$ and $b$ being replaced
by $N_{1(2)}$ and $b_{1(2)}$ for the incoming (outgoing) pion,
defined in Eq.~(\ref{choice of rapidity parameter}) diminishes the large logarithms
in the NLO TMD pion wave function in Eq.~(33) of \cite{LSW11}. It is indeed the case
under the power counting $k_{1T}^2 \sim k_{2T}^2 \sim x_1 \, x_2 \, Q^2$, confirming
the universality of a TMD hadron wave function. To eliminate the large logarithms
in the hard kernel given by Eq.~(35) of \cite{LSW11}, we may set
\begin{eqnarray}
\zeta^2_1= N_2^{-45/8} \, N_1^{1/2} \,,
\qquad \zeta^2_2= N_1^{-45/8} \, N_2^{1/2}\,,
\nonumber
\end{eqnarray}
which arise from the simultaneous solution to the evolution equations for two TMD
pion wave functions and one hard kernel.
We will present the details of the joint-resummation improved factorization
for the pion electromagnetic form factor elsewhere.

\section*{Acknowledgement}

We thank P. Masjuan and N.G. Stefanis for helpful comments on our paper.
HNL is supported  in part by the National Science Council of
R.O.C. under Grant No. NSC-101-2112-M-001-006-MY3, and by the National
Center for Theoretical Sciences of R.O.C..
YLS acknowledges the support of  National Science
Foundation of China under Grant No. 11005100.
YMW is supported by  the  DFG Sonderforschungsbereich
/Transregio 9 ``Computergest\"{u}tzte Theoretische Teilchenphysik".

\appendix

\section{Explicit expressions of the functions $F_i$}
\label{functions F}

The functions $F_i(\lambda_1,\lambda_2,\lambda_3,\lambda_4, \eta)$ ($i=1,2,3$)
appearing in the joint-resummation improved pion wave function
$\overline{\Phi}^{\rm (I,II,III)}(x,b,\zeta_1^2, t)$ in Eqs.~(\ref{ab1}),
(\ref{ab2}), and (\ref{ab3}) are defined as
\begin{eqnarray}
&& F_1(\lambda_1,\lambda_2,\lambda_3,\lambda_4, \eta)
\nonumber \\
&& = { C_F \over \beta_0}
\bigg \{ \hat {\lambda}_1 \, \left [  {1 \over 2} \ln \left( \hat {\lambda}_1 ^2
+{\pi^2 \over 4} \right) -1 \right ]  -{\pi \over 2} \theta_1(\lambda_1,\eta)
- \hat {\lambda}_2  \, \left [  {1 \over 2} \ln \left( \hat {\lambda}_2^2
+{9 \pi^2 \over 4} \right) -1 \right ] -{3 \pi \over 2} \theta_2(\lambda_2,\eta)
\nonumber \\
&& \hspace{0.5 cm}- \hat {\lambda}_3 \,  \left [  {1 \over 2} \ln \left(\hat {\lambda}_3^2
+{\pi^2 \over 4} \right) -1 \right ]  + { \pi \over 2} \theta_3(\lambda_3,\eta)
+ \hat {\lambda}_4 \,  \left [  {1 \over 2} \ln \left( \hat {\lambda}_4^2
+{9 \pi^2 \over 4} \right) -1 \right ]  + {3 \pi \over 2} \theta_4(\lambda_4,\eta)
\nonumber \\
&& \hspace{0.5 cm} -{1 \over 4} {\rm Li}_2 \left(-e^{-2  \, \hat {\lambda}_1 } \right)
+{1 \over 4} {\rm Li}_2 \left(-e^{-2 \, \hat {\lambda}_2} \right)
+{1 \over 4} {\rm Li}_2 \left(-e^{-2 \, \hat {\lambda}_3} \right)
-{1 \over 4} {\rm Li}_2 \left(-e^{-2 \, \hat {\lambda}_4} \right) \bigg \} \,,
\\
\nonumber \\
&& F_2(\lambda_1,\lambda_2,\lambda_3,\lambda_4, \eta)
\nonumber \\
&& ={ C_F \over \beta_0}
\bigg \{ \hat {\lambda}_1 \,  \theta_1(\lambda_1,\eta)
+ {\pi \over 4} \ln \left(\hat {\lambda}_1^2 +{\pi^2 \over 4} \right)
- \hat {\lambda}_2 \,  \theta_2(\lambda_2,\eta)
+ {3 \pi \over 4} \ln \left(\hat {\lambda}_2^2 +{9 \pi^2 \over 4} \right)
\nonumber \\
&& \hspace{0.5 cm} -  \hat {\lambda}_3 \,   \theta_3(\lambda_3,\eta)
- {\pi \over 4} \ln \left(\hat {\lambda}_3^2 +{\pi^2 \over 4} \right)
+ \hat {\lambda}_4 \,  \theta_4(\lambda_4,\eta) -{3 \pi \over 4}
\ln \left(\hat {\lambda}_4^2  +{9 \pi^2 \over 4} \right)
\nonumber \\
&& \hspace{0.5 cm} +{\rm  Im} \left [   {\rm Li}_2 \left( i e^{-  \hat {\lambda}_1 } \right)
-  {\rm Li}_2 \left( i e^{-  \hat {\lambda}_2} \right)
-   {\rm Li}_2 \left( i e^{-  \hat {\lambda}_3} \right)
+  {\rm Li}_2 \left( i e^{-  \hat {\lambda}_4 } \right)  \right ]
\bigg \} \,.
\\
\nonumber \\
&& F_3(\lambda_1,\lambda_2,\lambda_3,\lambda_4,\eta)
\nonumber \\
&&={ C_F \over \beta_0}  \bigg \{ \hat {\lambda}_1  \,
\bigg ( \ln \hat {\lambda}_1   -1  \bigg )
-\hat {\lambda}_2
\bigg ( \ln \hat {\lambda}_2  -1  \bigg )
- \hat {\lambda}_3
\bigg ( \ln \hat {\lambda}_3  -1  \bigg )
+ \hat {\lambda}_4 \,
\bigg ( \ln \hat {\lambda}_4 -1  \bigg )
\nonumber \\
&& \hspace{0.5  cm} -{\rm Li}_2\left(e^{-\hat {\lambda}_1  }\right)
+{\rm Li}_2\left(e^{-\hat {\lambda}_2  }\right)
+{\rm Li}_2\left(e^{-\hat {\lambda}_3  }\right)
-{\rm Li}_2\left(e^{-\hat {\lambda}_4  }\right) \bigg \} \,,
\end{eqnarray}
with the short-hand notations  $\hat {\lambda}_i$ and $\theta_i(\lambda_i,\eta)$
\begin{eqnarray}
\hat {\lambda}_{1(3)}&=\lambda_{1(3)} + {1 \over 2} \, \ln \eta \,,   \qquad
\hspace{2.3 cm} \hat {\lambda}_{2(4)}&=\lambda_{2(4)} - {3 \over 2} \, \ln \eta \,, \nonumber \\
\theta_1(\lambda_1,\eta)&= {\rm arctan} \left ({\pi  \over 2 \, \hat {\lambda}_{1}}  \right)
+\pi \theta \left ( - \hat {\lambda}_{1}  \right ) \,,  \qquad
\theta_2(\lambda_2,\eta)&= - {\rm arctan} \left ( {3 \, \pi  \over 2 \, \hat {\lambda}_{2} } \right )
-\pi \theta \left ( - \hat {\lambda}_{2}  \right ) \,,  \nonumber \\
\theta_3(\lambda_3,\eta)&= \theta_1(\lambda_3,\eta) \,,  \qquad
\hspace{2.5 cm} \theta_4(\lambda_4,\eta)&= \theta_2(\lambda_4,\eta) \,.  \nonumber
\end{eqnarray}

\end{document}